\documentclass[twocolumn]{aastex701}
\usepackage{amsmath}
\usepackage{multirow}
\usepackage{hyperref}
\usepackage[normalem]{ulem}
\usepackage{acronym}
\usepackage{xspace}

\acrodef{BSE}{binary stellar evolution}
\acrodef{BNS}{binary neutron star}
\acrodefplural{BNS}[BNSs]{binary neutron stars}
\acrodef{BBH}{binary black hole}
\acrodefplural{BBH}[BBHs]{binary black holes}
\acrodef{BHNS}{black hole--neutron star}
\acrodefplural{BHNS}[BHNSs]{black hole--neutron stars}
\acrodef{GSMF}{galaxy stellar mass function}
\acrodef{SMHM}{stellar-to-halo mass}
\acrodef{MZR}{mass-metallicity relation}
\acrodef{GW}{gravitational wave}
\acrodefplural{GW}[GWs]{gravitational waves}
\acrodef{DCO}{double compact object}
\acrodefplural{DCO}[DCOs]{double compact objects}
\acrodef{LVK}{LIGO-Virgo-KAGRA}
\acrodef{SNR}{signal-to-noise ratio}
\acrodef{FAR}{false-alarm rate}
\acrodef{BH}{black hole}
\acrodef{NS}{neutron star}
\acrodef{CE}{common envelope}
\acrodef{OSMT}{only stable mass transfer}
\acrodef{SFR}{star formation rate}
\acrodef{SMT}{stable mass transfer}
\acrodef{SN}{supernova}
\acrodefplural{SN}[SNe]{supernovae}
\acrodef{PISN}{pair-instability supernova}
\acrodefplural{PISN}[PISNe]{pair-instability supernovae}
\acrodef{WD}{white dwarf}
\acrodefplural{WD}[WDs]{white dwarfs}
\acrodef{ZAMS}{zero-age main sequence}

\DeclareRobustCommand{\VAN}[3]{#2}
\let\VANthebibliography\thebibliography
\def\thebibliography{\DeclareRobustCommand{\VAN}[3]{##3}\VANthebibliography}
\usepackage{mathtools}	
\usepackage{tabularx}
\usepackage{color}
\usepackage{enumitem}
\usepackage{wrapfig}
\usepackage{outlines}

\usepackage{acronym}   
\usepackage{xspace}

\acrodef{GSMF}{galaxy mass function, the number density of galaxies per logarithmic mass bin}
\acrodef{MZR}{mass-metallicity relation}
\acrodef{SFRD}{star formation rate density}

\acrodef{BNS}{binary neutron star}
\acrodef{BBH}{binary black hole}
\acrodef{BHNS}{black hole - neutron star}

\acrodef{DCO}{double compact object}
\acrodef{NS}{neutron star}
\acrodef{BH}{black hole}
\acrodef{BH--NS}{black hole-neutron star}
\acrodef{GRB}{gamma--ray burst}
\acrodef{RLOF}{Roche-lobe overflow}
\acrodef{CE}{common envelope}

\newcommand{\Msun}{\ensuremath{\,\rm{M}_{\odot}}\xspace}

\newcommand{\Zsun}{\ensuremath{\,\rm{Z}_{\odot}}\xspace}

\newcommand{\SFRD}{\ensuremath{\mathcal{S}(Z,z)}\xspace}

\begin{document}

\title{From cosmological simulations to binary black hole mergers: \\ The impact of using analytical star formation history models on gravitational-wave source populations}

\correspondingauthor{Sasha Levina}

\author[0000-0003-1241-7615]{Sasha Levina}
\affiliation{Department of Astronomy \& Astrophysics, University of California, San Diego, 9500 Gilman Drive, La Jolla, CA 92093, USA}
\affiliation{William H. Miller III Department of Physics and Astronomy, Johns Hopkins University, 3400 N. Charles Street, \\ Baltimore, Maryland, 21218, USA}
\email[show]{slevina@ucsd.edu}

\author[0000-0002-4421-4962]{Floor Broekgaarden}
\affiliation{Department of Astronomy \& Astrophysics, University of California, San Diego, 9500 Gilman Drive, La Jolla, CA 92093, USA}
\email{fbroekgaarden@ucsd.edu}

\author[0000-0001-5484-4987]{Lieke van Son}
\affiliation{Radboud University, Faculty of Science, Huygensgebouw, Heyendaalseweg 135, 6525 AJ Nijmegen, The Netherlands}
\email{lieke.vanson@ru.nl}

\author[0000-0003-0751-5130]{Emanuele Berti}
\affiliation{William H. Miller III Department of Physics and Astronomy, Johns Hopkins University, 3400 N. Charles Street, \\ Baltimore, Maryland, 21218, USA}
\email{berti@jhu.edu}

\author[0000-0001-9583-4339]{Amedeo Romagnolo}
\affiliation{Department of Astronomy \& Astrophysics, University of California, San Diego, 9500 Gilman Drive, La Jolla, CA 92093, USA}
\affiliation{Nicolaus Copernicus Astronomical Center, Polish Academy of Sciences, ul. Bartycka 18, 00-716 Warsaw, Poland}
\affiliation{Universit\"at Heidelberg, Zentrum f\"ur Astronomie (ZAH), Institut f\"ur Theoretische Astrophysik, Albert Ueberle Str. 2, 69120, Heidelberg, Germany}
\affiliation{Dipartimento di Fisica e Astronomia Galileo Galilei, Università di Padova, Vicolo dell’Osservatorio 3, I–35122 Padova, Italy}
\email{amedeo.romagnolo@uni-heidelberg.de}

\author[0000-0003-3308-2420]{Ruediger Pakmor}
\affiliation{Max Planck Institute for Astrophysics, Karl-Schwarzschild-Str. 1, D-85748 Garching, Germany}
\email{rpakmor@mpa-garching.mpg.de}

\author[0009-0003-9165-9889]{Ana Lam}
\affiliation{Department of Astronomy, Columbia University, 538 West 120th Street, New York, NY 10027, USA}
\email{ana.lam@columbia.edu}

\begin{abstract}

Observations of \ac{BBH} mergers provide a unique window into the lives of massive stars across cosmic time.  Connecting redshift-dependent merger properties to massive star progenitors requires accurate models of cosmic star formation and chemical enrichment histories. Analytical fits for the metallicity-specific cosmic star formation rate density \SFRD are commonly used as proxies for the complex underlying star formation history, yet they remain unconstrained. Using the IllustrisTNG cosmological simulations, we evaluate the accuracy of these analytical \SFRD prescriptions. By coupling the simulated and analytical \SFRD to the population synthesis code {\tt COMPAS}, we investigate the resulting \ac{BBH} merger rates and mass distributions.  We find that analytical \SFRD prescriptions can overestimate \ac{BBH} merger rates at high redshift ($z\gtrsim6$) by up to a factor of $10$--$10^4$, depending on cosmological simulation resolution, and can introduce spurious features in the \ac{BBH} mass distribution.  For example, they can produce an artificial feature near $8\,M_\odot$ in the primary mass distribution at $z\lesssim2$, which is absent when using the full simulation-based \SFRD, while simultaneously suppressing high-mass features. These discrepancies arise because simple analytical models fail to capture a high-metallicity bump and a more flattened low-metallicity tail in the simulated \SFRD metallicity distribution. Our results highlight the importance of accurate star formation histories for modeling \ac{BBH} populations, demonstrate the limitation of widely used  analytical \SFRD fits, and underscore the need for careful integration of cosmological simulations, analytical fits, and population synthesis when interpreting gravitational-wave observations.

\end{abstract}


\section{Introduction}
\label{sec:intro}


%
%
The mergers of \Acp{BBH}, even those occurring nearby, encode information about stellar physics across a large range of redshifts due to the distribution of delay times between their formation and merger \citep{Fishbach2021, MandelFarmer2022, Fishbach2023, Chruslinska2024}. To disentangle the mixture of delay times and formation channels within an observed \ac{GW} population and to infer the environments and binary evolutionary processes that produce merging \acp{BBH}, astrophysical population synthesis models of their formation pathways are essential \citep{Broekgaarden2021, Broekgaarden2022, MandelBroekgaarden2022}. These models must be combined with assumptions about the cosmic star formation history in order to accurately represent stellar progenitors across cosmic epochs \citep[e.g.,][]{Neijssel2019, Chruslinska2019a, Broekgaarden2022, Boesky2024}. 

In modeling rates of metallicity-dependent transients (e.g., \ac{BBH} mergers), the star formation history is commonly represented using a simple, fitted functional form that combines a star formation rate density $\mathcal{S}(z)$ with some assumption for the metallicity distribution across redshift \citep[e.g.,][]{Neijssel2019}. The resulting metallicity-dependent star formation rate density \SFRD can vary significantly depending on the model, how it is constructed, and the data used to calibrate its parameters \citep[see][]{Chruslinska2024, Chruslinska2025}. Rates of double compact object mergers across redshift may change by an order of magnitude depending on the assumptions made for \SFRD \citep{Chruslinska2019a, Neijssel2019, Santoliquido2021, Broekgaarden2022, vanSon2023}. 
Testing the effect of different models of \SFRD can help disentangle the effects of metallicity evolution, the \ac{SFR}, and binary evolution models on the \ac{BBH} population and provide insight into which components are most critical to model accurately.

Three methods are typically used to model the star formation history: (i) analytical prescriptions, (ii) observational distributions, or (iii) cosmological simulations. Analytical prescriptions for $\mathcal{S}(z)$ and the redshift-dependent metallicity distribution \citep[as employed by e.g.,][]{LangerNorman2006, MadauDickinson2014, MadauFragos2017, Neijssel2019, Tang2020, Santoliquido2020, Boesky2024, Smith2024, Sgalletta2024, Turbang2024, Schiebelbein2024} are derived from galaxy survey data or empirical scaling relations such as $\mathcal{S}(z)$, the \ac{MZR}, and the \ac{GSMF}.
Observational data can also be directly used to create a distribution of star formation histories without performing an analytical fit \citep{Chruslinska2019a, Chruslinska2019b, Boco2021, Sgalletta2024}. However, observations become increasingly challenging at high redshifts and low galaxy luminosities, with uncertainties in high-redshift dust extinction, survey sensitivity and sky coverage limits, and discrepancies between star formation histories derived from different observational methods \citep{Chruslinska2019b,Katsianis2020,Katsianis2021,Enia2022,Magnelli2024}. This makes it more challenging to construct a comprehensive cosmic star formation history.

Cosmological simulations offer a third approach by modeling a representative population of galaxies across cosmic history. These simulations track the characteristics of stellar populations and the gas from which they form over redshift \citep[for a review, see][]{Vogelsberger2020}. This enables the construction of the complete \SFRD from the \ac{SFR} and metallicity of galaxies, accounting for every galaxy formed in the simulation box without observational limitations \citep[e.g.,][]{Hwang2019, Briel2022}. The accuracy of that star formation history is constrained by the model resolution, physical assumptions built into the simulation, and the treatment of unresolved processes via subgrid models \citep{Pillepich2018b, Vogelsberger2020, Pakmor2022}. Different simulations can deviate particularly at high redshifts, as high redshift galaxy evolution is not directly accounted for in the model calibration. The resulting \ac{BBH} merger rate can vary significantly with simulation used \citep{Briel2022}.  
Others use the cosmological simulation \SFRD implicitly by populating \acp{BBH} (or other transients) into the simulation by associating them with stellar populations in the simulation \citep[e.g.,][]{Mapelli2017, Mapelli2018, Mapelli2019, Artale2019, Artale2020, Marinacci2025}. More recently, analytical fits to cosmological simulations have been used to represent \SFRD \citep{vanSon2023}, which is computationally simpler for further use with population synthesis than constructing \SFRD from the simulation data. These fits have become widely used \citep[e.g.][]{vanSon2022b, Hendricks2023, Riley2023, Roy2025, Willcox2025}.

The main challenge in modeling \ac{BBH} populations using analytical fits to cosmological simulations is twofold.
First, it is unclear whether analytical prescriptions can reliably represent the underlying \SFRD.
Second, the impact of both \SFRD assumptions and numerical effects (differences across cosmological simulations based on factors not related to the astrophysical or cosmological assumptions e.g., mass resolution or box volume) on the star formation history, and hence on the predicted \ac{BBH} population, remains insufficiently understood.
Several studies have addressed parts of this problem.
\citet{vanSon2023} tested the goodness of fit of analytical \SFRD models to a single IllustrisTNG simulation and examined how varying \SFRD parameters affects the \ac{BBH} primary mass distribution.
\citet{Briel2022} explored how different cosmological simulation suites influence predicted \ac{BBH} population properties.
\citet{Sgalletta2024} compared analytical \SFRD prescriptions with observationally derived models when computing the \ac{BBH} merger rate.
With respect to numerical effects, \citet{Marinacci2025} found that variations in box size have a negligible effect on \SFRD and on the resulting \ac{BBH} properties.
By contrast, resolution effects are known to systematically affect \SFRD. Higher-resolution simulations tend to yield higher star formation rate normalizations \citep{Pillepich2018b}.
Some simulation suites mitigate this by recalibrating their subgrid models at each resolution \citep[see][]{Marinacci2014}.
However, the consequences of resolution variations for modeling \ac{BBH} populations have not yet been studied in a systematic way.
A quantitative assessment of both analytical modeling choices and numerical effects is therefore essential. An accurate description of the cosmic star formation history is a prerequisite for robust predictions of the \ac{BBH} merger population. This motivates the analysis we present in this work.

In this paper, we quantify the impacts of approximating the star formation history of cosmological simulations using analytical models and investigate the effect of cosmological simulation resolution on the \ac{BBH} merger population. We fit analytical \SFRD models to three cosmological simulations of varying resolution and box size in the IllustrisTNG suite, and compare the resulting star formation histories and \ac{BBH} population properties using the binary population synthesis code {\tt COMPAS} \citep[Compact Object Mergers: Population Astrophysics and Statistics,][]{COMPAS}.

\section{Methods}
\label{sec:methods}

The metallicity-dependent star formation rate density \SFRD quantifies the amount of stellar mass formed per unit comoving volume, redshift, and metallicity.
In this study, we construct \SFRD in two ways: (i) directly based on the IllustrisTNG simulations (see Section~\ref{sec:cosmological sim}), and (ii) using an analytical fit that is often used in the field to represent \SFRD for the TNG simulation (see Section~\ref{sec:model}).
We use both in the convolution (cosmic integration, see Section~\ref{sec:compas}) to model the \ac{BBH} merger population and investigate how using an analytical fit causes deviations from directly using the underlying distribution from the TNG simulations with different resolutions and simulation box volumes (see Section~\ref{sec:comparison methods}). 

\subsection{Obtaining the TNG simulation \SFRD}
\label{sec:cosmological sim}

We derive the simulation-based \SFRD from the IllustrisTNG simulations \citep{Nelson2019a, Pillepich2018a, Marinacci2018, Naiman2018, Nelson2018, Nelson2019b, Pillepich2019}.
IllustrisTNG is a suite of magnetohydrodynamical cosmological simulations performed with the moving-mesh code \textsc{AREPO} \citep{arepo} and adopting cosmological parameters $\Omega_M =0.3089, \Omega_\Lambda=0.691, \sigma_8=0.8159, n_s=0.9667, h=0.6774, H_0=67.8$ \citep{planck2016}.

The IllustrisTNG simulations span three primary volumes: TNG50-1, TNG100-1, and TNG300-1, with comoving box volumes with side lengths of approximately $50$, $100$, and $300$ Mpc and baryonic mass resolutions of $5.7 \times 10^4$, $9.4 \times 10^5$, and $7.6\times 10^6 \Msun/h$, respectively. 
The highest resolution box, TNG50-1, resolves the internal structure of galaxies, including stellar populations and gas inflows and outflows. 
On the other hand, TNG300-1, while lower in resolution, captures a representative population of galaxy clusters and provides advantages through providing a larger statistical sample, but does not form low-mass galaxies below its minimum resolution. 
TNG100-1 offers an intermediate balance between resolution and volume, it is directly comparable to the predecessor Illustris simulations \citep{Nelson2019a}, and it is more similar to other cosmological simulations. 
In the following, we focus on the highest-resolution realizations of each box (TNG50-1, TNG100-1, and TNG300-1).



Each TNG simulation consists of data output at 100 redshift snapshots from $z=20$ to $z\approx0$. 
Snapshots include the full set of resolution elements (gas, dark matter, stars, etc.) for each cells in the box at that redshift snapshot, divided into smaller ``chunks'' for data handling. 
For our analysis we only use the gas cells (``PartType0''), as they contain the metallicity and instantaneous \ac{SFR} information needed to compute \SFRD. 

Star formation occurs in cells with gas densities exceeding  $n_\text{H} \gtrsim 0.1 \ \text{cm}^{-3}$, following the Kennicutt-Schmidt relation prescription \citep{Schmidt1959, Kennicutt1989}. Star formation occurs stochastically and typically converts a cell fully into a single star particle. If that happens, the star particle represents an unresolved stellar population with the mass of the star forming cell, which is within a factor of two of the target baryonic mass resolution of the simulation \citep{Pakmor2022}.
The metallicity of a gas cell, defined as the mass fraction of all elements heavier than helium, characterizes the metallicity of the associated stellar population. 

To construct the simulation \SFRD, we bin all star-forming gas cells in each snapshot by metallicity. 
We use 60 logarithmically-spaced metallicity bins between the minimum TNG metallicity of $\log_{10}Z=-10$ and the maximum $\log_{10}Z=0$, summing the \ac{SFR} of all gas cells in each bin. Note that the choice of binning here does not affect the final results. 
The resulting \SFRD is normalized by the comoving simulation volume to yield units of $M_\odot\,\rm{yr}^{-1}\,\rm{Mpc}^{-3}$. 
We also retrieve the overall gas metallicity distribution through binning per snapshot, $\mathrm{dP/d}\log{Z}$, and record this for later comparison.

To ensure smooth coverage across metallicity and redshift, we do a linear interpolation on the simulation \SFRD over the range $0 < z < 14$ with steps of $\Delta z = 0.05$ and 500 logarithmic metallicity bins. The interpolation is performed in lookback time (rather than redshift) using the \cite{planck2016} cosmology, consistent with that adopted in IllustrisTNG.

\subsection{Modeling the metallicity-dependent star formation rate density with an analytical fit}
\label{sec:model}


The metallicity-dependent star formation rate density \SFRD is often modeled as the product of the total star formation rate density and a metallicity distribution \citep[e.g.,][]{Dominik2013, Chruslinska2019a, Chruslinska2019b, Neijssel2019, 
Broekgaarden2021, vanSon2023}.
This product is represented by using an analytical fit because it is simple, fast, convenient for uncertainty quantification, and parametrizes the distribution into physical variables \citep{Neijssel2019, vanSon2023, Willcox2025}. 
We investigate whether such analytical fits are a reliable approximation for \SFRD by fitting the \citet{vanSon2023} \SFRD prescription to the TNG simulations.
The analytical \SFRD is constructed by combining the total star formation rate density, $\mathcal{S}(z)$, with a skewed log-normal metallicity distribution, dP/dZ$(Z, z)$:
\begin{equation}
    \mathcal{S}(Z,z) = \mathcal{S}(z) \times \frac{\text{dP}}{\text{dZ}}(Z, z).
\label{eq:sfrd(Z,z)}
\end{equation}

The quantity $\mathcal{S}(z)$ is the total mass formed, per comoving volume per year. 
It is defined using the functional form of \citet{MadauDickinson2014}:

\begin{equation}
    \mathcal{S}(z) = a\frac{(1+z)^b}{1+[(1+z)/c]^d},
\label{eq:sfr}
\end{equation}
where $a$ is a normalization factor, $b$ characterizes the low redshift slope (where $\mathcal{S}(z)$ scales as $(1+z)^b$ away from the peak at low $z$), $c$ defines the location of the peak, and $d$ dominates the high redshift slope \citep[see Appendix A.1 of][]{Iacovelli2022}. 
The \citet{MadauFragos2017} values for $\mathcal{S}(z)$ parameters are $a=0.01$, $b=2.6$, $c=3.2$, and $d=6.2$. 

For the metallicity distribution function $\text{dP/dZ}(Z, z)$, we adopt a skewed log-normal distribution in metallicity and redshift following \citet{vanSon2023}.
We choose to use a skewed log-normal to represent the metallicity distribution as opposed to a log-normal \citep[as has been used in e.g.,][]{Belczynski2016, Neijssel2019} as a number of recent works have shown that the metallicity distribution is not symmetrical and should have a tail towards low metallicities---primarily attributed to low mass and starburst galaxies---by modeling the metallicity distribution and \SFRD derived from observations \citep{Chruslinska2019b, Boco2021, Chruslinska2021, Santoliquido2022, Sgalletta2024, Boco2026}, as well as cosmological simulations \citep{Pakmor2022, vanSon2023}.
We therefore model the metallicity distribution as
\begin{equation}
    \frac{\text{dP}}{\text{dZ}}(Z, z) = \frac{2}{\omega(z) Z} \times \phi\left(\frac{\ln{Z} - \xi(z)}{\omega(z)} \right) \Phi \left( \alpha \frac{\ln{Z} - \xi(z)}{\omega(z)} \right),
\label{eq:dP/dZ}
\end{equation}
where $\phi$ is the standard log-normal distribution,
\begin{equation}
  \phi \left ( \frac{\ln{Z} - \xi}{\omega} \right ) = \frac{1}{\sqrt{2 \pi}} \exp \left[ - \frac{1}{2} \left (\frac{\ln{Z} - \xi}{\omega} \right ) ^2 \right];
\end{equation}
and $\Phi$ is its cumulative distribution function,
\begin{equation}
  \Phi \left (\alpha\frac{\ln{Z} - \xi}{\omega} \right ) = \frac{1}{2} \left[ 1 + \text{erf} \left (\alpha\frac{\ln{Z} - \xi}{\omega\sqrt{2}} \right ) \right].
\end{equation}
The ``skewness'' parameter $\alpha$ introduces asymmetry into the distribution. The function $\xi(z)$ determines the median of the distribution and is given by  
\begin{equation}
    \xi(z) = \ln \left [ \frac{\mu_0 \, 10^{\mu_z \, z}}{2 \Phi ({\alpha \omega/\sqrt{1+\alpha^2}})} \right ] - \frac{\omega^2}{2}.
\label{eq:xi}
\end{equation}

The metallicity distribution function dP/dZ$(Z, z)$ is characterized by five physically motivated parameters: $\mu_0, \mu_z, \omega_0, \omega_z$, and $\alpha$.  
The parameter $\mu_0$ sets the mean metallicity at $z\approx0$, while $\mu_z$ describes its redshift evolution according to $\mu(z) = \mu_0 \, 10^{\mu_z \, z}$ \citep[e.g.][]{LangerNorman2006,Neijssel2019,vanSon2023}. 
The width of the metallicity distribution is represented by $\omega_0$ at $z\approx0$ and evolves across redshift as $\omega(z) = \omega_0 \,10^{\omega_z \, z}$.
The final parameter, $\alpha$, encodes the asymmetry of the metallicity distribution:  $\alpha=0$ corresponds to a symmetric log-normal distribution, while $\alpha < 0$ produces a longer low-metallicity tail. 
Following \citet{vanSon2023}, we assume that $\alpha$ does not evolve with redshift; see \citet{vanSon2023} for a more detailed discussion of this parameterization.
Together, the parameters describing $\mathcal{S}(z)$ and dP/dZ$(Z, z)$ define a nine-parameter analytical fit that is often used to represent \SFRD. We will be using this \SFRD for the analytical fit to the TNG simulations.

Finally, we fit the analytical model described above to the interpolated simulation \SFRD described in Section~\ref{sec:cosmological sim} using the optimization framework of \citet{vanSon2023} to obtain the best-fit analytical \SFRD to each TNG. 
This optimization employs the cost function described in \citet{vanSon2023}, designed to optimize for the fit at both the peak of \SFRD and the low-metallicity tail important for the formation of merging \ac{BBH} systems. Therefore, a penalization factor is applied to reduce the contribution of the star formation peak to the fit. This prevents the optimization from being dominated by the high values of \SFRD near $z\sim 2$.

We employ the {\tt SciPy~v1.8.1} implementations of the {\tt BFGS} optimization algorithm \citep{Nocedal2006} for TNG50-1 and TNG100-1, and the {\tt Nelder-Mead} method \citep{Gao2012} for TNG300-1, as the latter shows improved convergence for lower-resolution data.


\subsection{Calculating BBH merger rates and properties over redshift with COMPAS}
\label{sec:compas}

{\tt COMPAS} \citep{COMPAS} is a rapid binary population synthesis code that is based on the stellar and binary evolution prescriptions of \citet{Hurley2000, Hurley2002}. 
We use {\tt COMPAS} version v02.26.03 and adopt the simulations from \citet{vanSon2022a, vanSon2022b},  publicly available at \citet{ZenodoCOMPASrun}. 
The simulations contain $10^7$ binaries, selected using the adaptive importance sampling for binary black hole mergers STROOPWAFEL \citep{Broekgaarden2019}. 
Primary masses ($M_1$) were drawn from the \citet{Kroupa2001} initial mass function over the range  $10 \leq M_1/M_\odot \leq 150$. 
Secondary masses ($M_2$) were selected from a uniform mass ratio distribution, $q = M_2/M_1$, with $0.01 \leq q \leq 1.0$ and $M_2 \geq 0.1\,M_\odot$. 
Birth metallicities were drawn from a flat distribution in log-space between $10^{-4} \leq Z \leq 0.03$. 
The initial orbital separation follows a flat-in-log distribution in the range from $0.01$ to $1000$~AU. All binaries are assumed to be born circular at birth, consistent with observations that OB stars (likely progenitors of \acp{BBH}) tend to have circular orbits \citep{Vargas-Salazar2025}. 
Mass transfer events would further circularize the binary \citep[as is implemented in \texttt{COMPAS} following, e.g.,][]{Hurley2002, Belczynski2002, Belczynski2008}. 
It is important to note that it has been recently shown that a substantial fraction of binaries may retain eccentricity following an eccentric mass transfer event \citep{Rocha2025}.
For additional details on the {\tt COMPAS} simulation, see  \citet{vanSon2022a}. 

\subsection{Comparing the analytical fit and simulation \SFRD method}
\label{sec:comparison methods}

To study the impact of using the analytical fit for \SFRD, we compare the analytical fit and simulation \SFRD through two different methods. First, we compare \SFRD that the two methods produce. Second, we also investigate the \ac{BBH} properties (rates and mass distributions across redshift) that each method results in and compare this to observations. 

For comparing the \ac{BBH} properties, we convert the synthetic binary population from {\tt COMPAS} into a cosmic merger rate density for \ac{BBH} binaries by combining the binary formation efficiencies and \ac{BBH} properties from {\tt COMPAS} with either the analytical fit or the TNG simulation \SFRD model using the {\tt COMPAS} code. 
The calculation of \ac{BBH} merger rates and properties across redshift using ``cosmic integration'' post-processing scripts is based on the framework developed by \citet{Neijssel2019}, enabling flexible modeling of \SFRD through the parametrized model described in Section~\ref{sec:model}. 

To incorporate \SFRD directly extracted from the cosmological simulations, we implement a new method of modeling \SFRD in the {\tt COMPAS} cosmic integration framework that constructs \SFRD from the TNG simulations (see Section~\ref{sec:cosmological sim}). 
This method takes \SFRD as a two-dimensional array as a function of redshift and metallicity, along with the corresponding lookback times, redshifts, metallicity bin edges, and the binned metallicity distributions (dP/dlogZ) at each redshift. We then calculate the \ac{BBH} properties across redshift following \citet{Neijssel2019, vanSon2023, COMPAS}. Our code is publicly available at \citet{Zenodo}.
For the analytical fit, we perform the same method using the analytical fit \SFRD described in Section~\ref{sec:model}, which was already implemented in the cosmic integration framework by \citet{vanSon2023}.

Once the merger rate weights have been computed, \texttt{COMPAS} provides access to the distributions of binary properties, such as masses, spins, formation redshifts, and formation metallicities of merging compact objects at any given redshift, which we will discuss in our results and compare to the GWTC-4 results from \ac{LVK} \citep{LVK2025}.
To study different evolutionary pathways, we classify binaries that experienced at least one common-envelope (CE) phase as CE-channel systems, and binaries that avoided CE evolution but underwent mass transfer as stable-channel systems.

We assess the accuracy of the analytical fit \SFRD by comparing it directly to \SFRD obtained from the TNG simulations, as well as by comparing the resulting \ac{BBH} populations created with each \SFRD method.   
Specifically, throughout our work we compute the fit-to-simulation ratio for \SFRD,
\begin{align}
    \frac{\mathrm{fit} \ \mathcal{S}(Z, z)}{\mathrm{simulation} \ \mathcal{S}(Z, z)},
\end{align}
which quantifies where the fit overestimates or underestimates the simulated \SFRD, 
as well as the fit-to-simulation ratio for the \ac{BBH} merger properties,
\begin{align}
    \frac{\mathrm{fit}  \, d\mathcal{R}(z)/dx }{\mathrm{simulation} \,  d\mathcal{R}(z)/dx},
\end{align}
which quantifies where the fit overestimates or underestimates the simulated \ac{BBH} distribution (where $dx$ can represent redshift $dz$ or primary mass $dM_\mathrm{BH,1}$). 
We use this ratio rather than an absolute difference because the relevant quantities—such as \SFRD, merger rate, and mass distribution—span several orders of magnitude. The ratio representation therefore provides a clearer  measure of the fit quality across the parameter of interest.

\section{Results}
\label{sec:results}

\begin{table}[t]
    \centering
    \caption{Best fit \SFRD parameters.}
    \begin{tabular}{c c c c c}
    \hline
        & & TNG50-1 & TNG100-1 & TNG300-1 \\
        \hline
        \multirow{6}{*}[-0.4ex]{\rotatebox{90}{dP/dlogZ}} &&&&\\
        &$\mu_0$   & 0.0282  & 0.0247 & 0.0237 \\
        &$\mu_z$   & -0.0314  & -0.0521  & -0.0687 \\
        &$\omega_0$   &  1.1136 & 1.1509 &  1.1196 \\
        &$\omega_z$   & 0.0592 & 0.0477 &  0.0481 \\
        &$\alpha$   & -1.7353 & -1.8801 &  -2.2726 \\
        \multirow{5}{*}[-0.4ex]{\rotatebox{90}{$\mathcal{S}(z)$}} &&&&\\
        &$a$   & 0.0329 & 0.0172 &  0.0097\\
        &$b$   & 1.4668 & 1.4425 &  1.5747\\
        &$c$   & 3.8412 & 4.5299 &  4.5428\\
        &$d$   & 5.0994 & 6.2261 &  6.8266\\
    \hline
    \end{tabular}
    \label{tab:sfrd params}
\end{table}

\begin{figure*}[t]
    \centering
    \includegraphics[width=0.7\textwidth,angle=0]{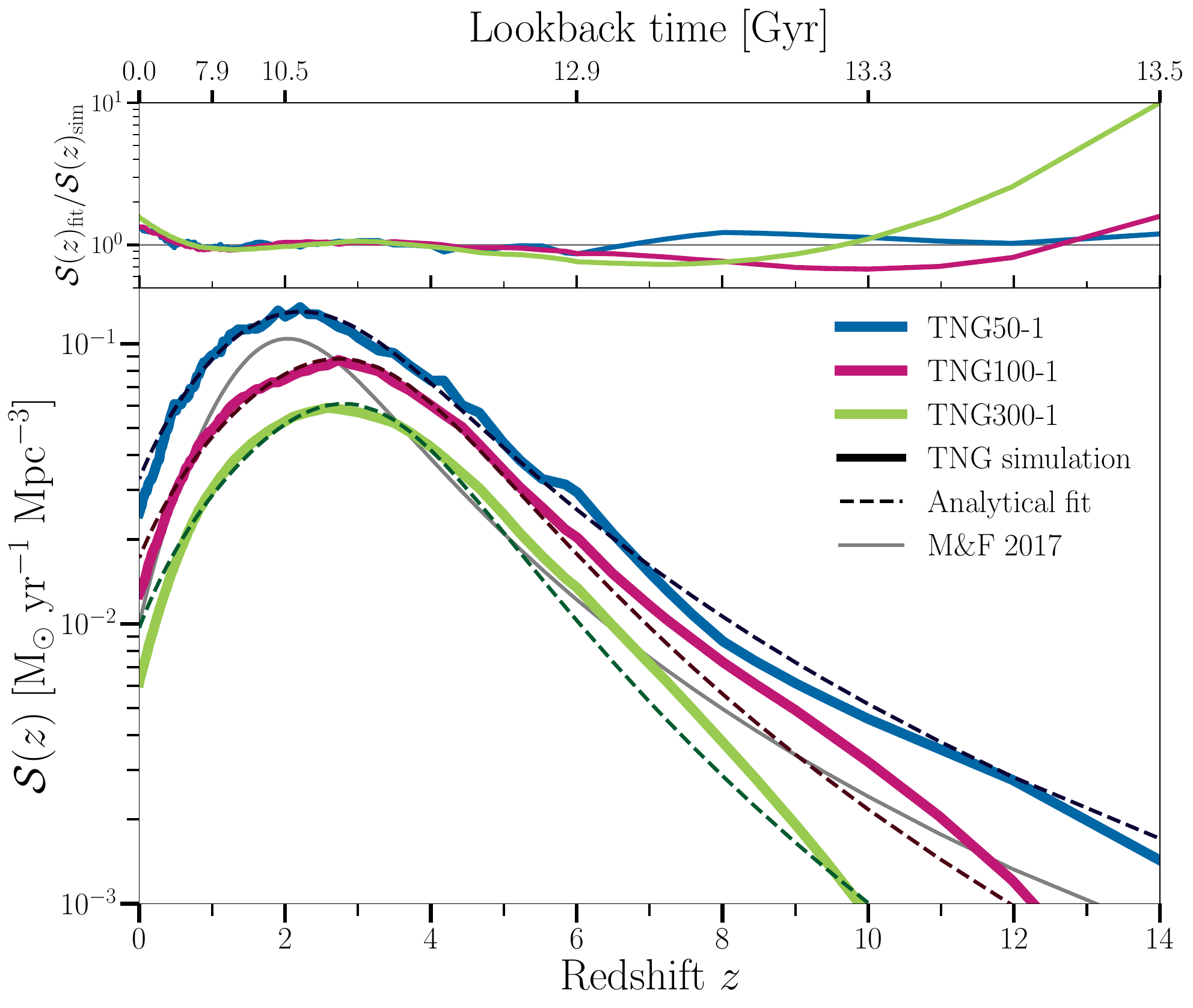}
    \caption{\textbf{Bottom panel:} Star formation rate density, $\mathcal{S}(z)$, as a function of redshift, $z$, for the IllustrisTNG simulations TNG50-1 (blue), TNG100-1 (magenta), and TNG300-1 (green). Solid lines show $\mathcal{S}(z)$ obtained by summing the full TNG \SFRD from each simulation over metallicity, while dashed lines are analytical fits to the TNG \SFRD. For comparison, $\mathcal{S}(z)$ model from \citet{MadauFragos2017}, which has the parameters  $a=0.01$, $b=2.6$, $c=3.2$, and $d=6.2$ as defined in Equation~\eqref{eq:sfr}, is shown as a gray line. \textbf{Top panel:} The fractional error for $\mathcal{S}(z)$ as a function of redshift, as described in Section~\ref{sec:comparison methods}.}
    \label{fig:sfr_z}
\end{figure*}

\begin{figure*}[t]
    \centering
    \includegraphics[width=0.7\textwidth,angle=0]{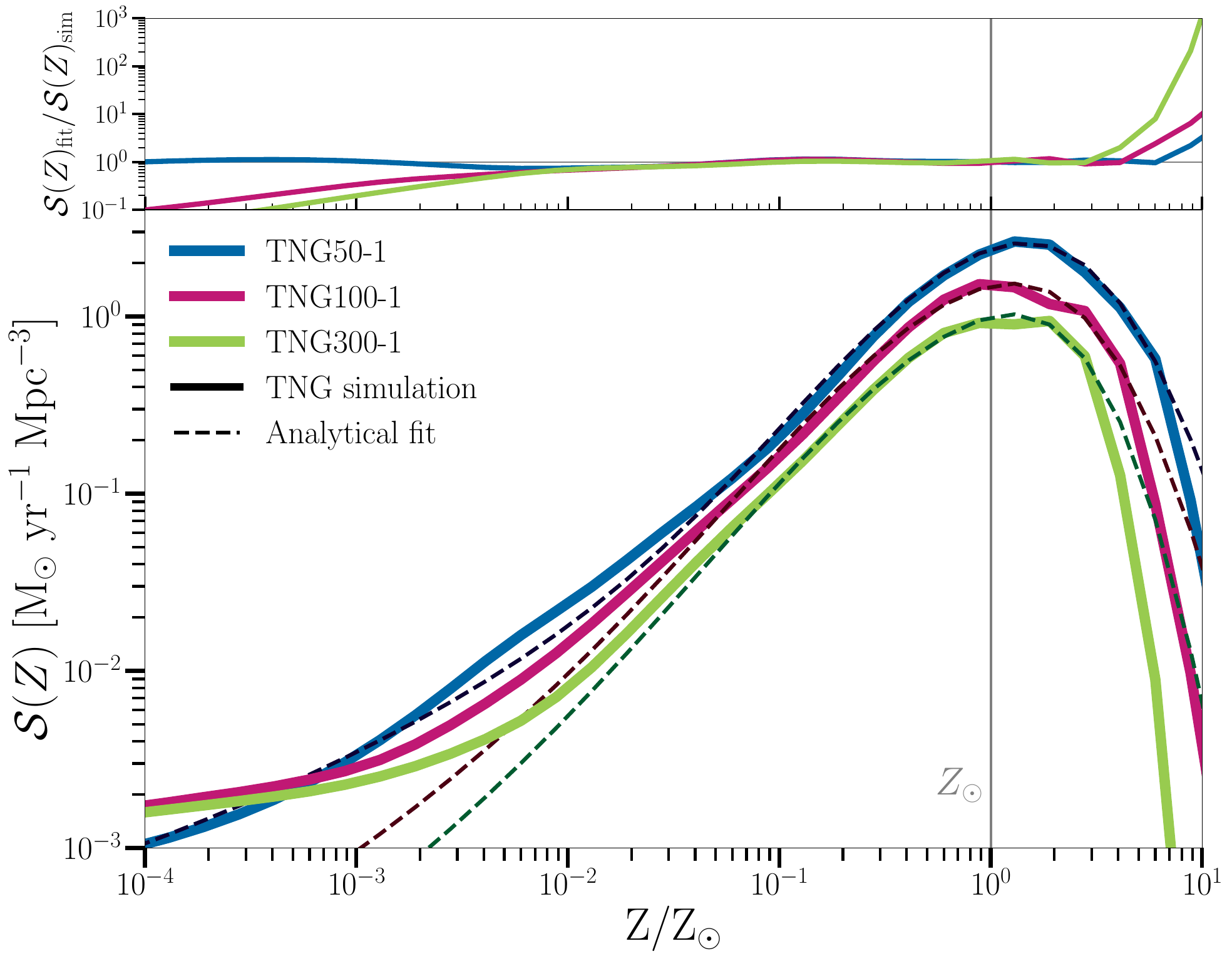}
    \caption{\textbf{Bottom panel:} Star formation rate density, $\mathcal{S}(Z)$, as a function of $Z/Z_\odot$. The colors and line styles are analogous to Figure \ref{fig:sfr_z}. \textbf{Top panel:} The percent error for $\mathcal{S}(Z)$.}
    \label{fig:sfr_Z}
\end{figure*}

\begin{figure*}
    \centering
    \includegraphics[width=1.0\textwidth,angle=0]{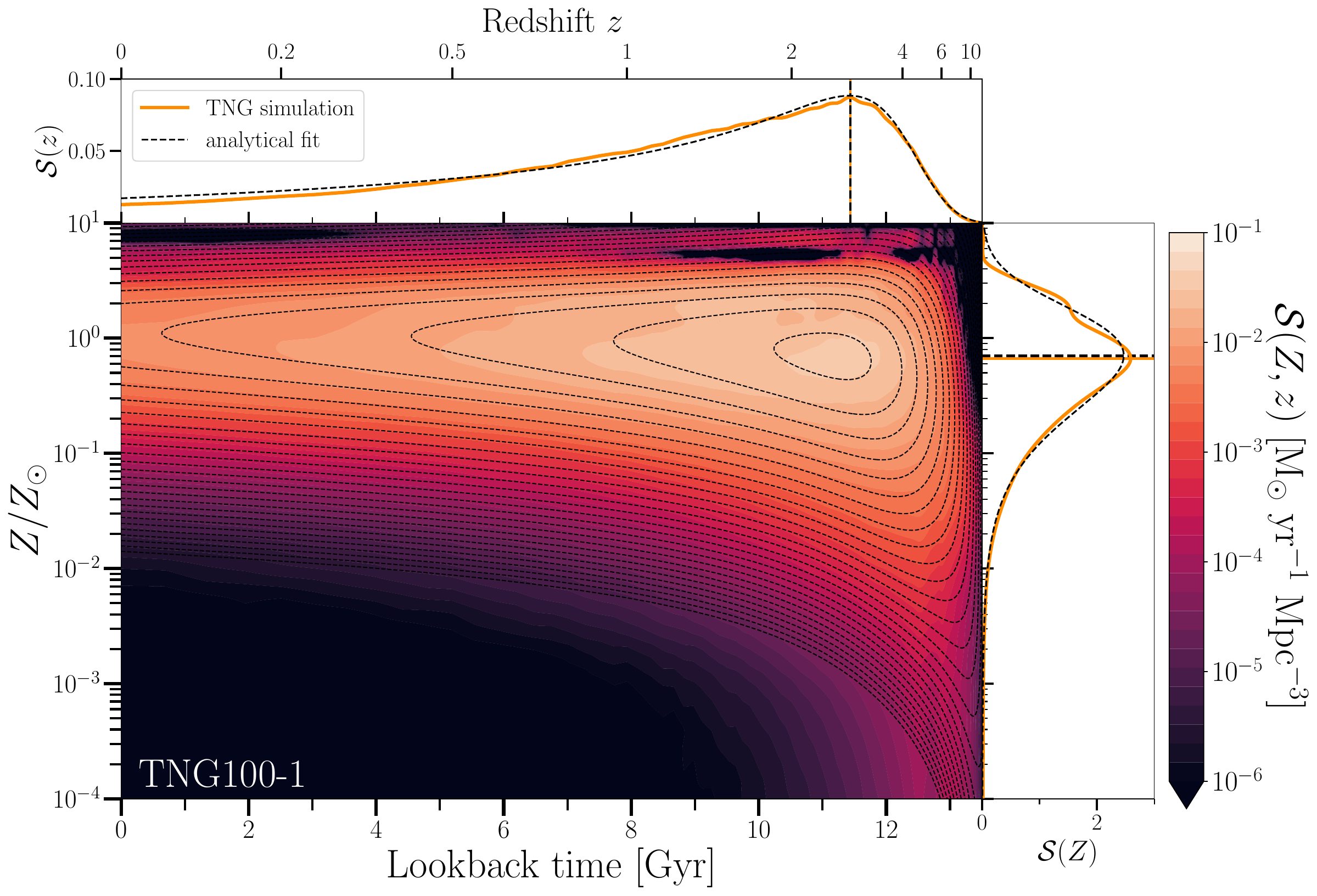}
    \caption{TNG100-1 \SFRD as a function of lookback time (Gyr) and $Z/Z_\odot$. The color bar shows the simulation \SFRD and the black dashed contours show \SFRD from the analytical fit. The top panel shows \SFRD marginalized over metallicity, and the right panel over lookback time. Note that the side panels are plotted using a linear scale for $\mathcal{S}(z)$ and $\mathcal{S}(Z)$. Black regions near the upper right reflect numerical noise from the interpolation edge.}
    \label{fig:SFRD_Z_z_tng100}
\end{figure*}

\subsection{Fit vs. simulation: Impact on \SFRD}
\label{sec:SFRD results}

The best-fit parameters for the analytical \SFRD model to \SFRD extracted from the TNG simulations (see Section~\ref{sec:model}) are listed in Table~\ref{tab:sfrd params}. Figure~\ref{fig:sfr_z} compares the simulation $\mathcal{S}(z)$ to the corresponding analytical fits for the three TNGs used. 
Here, the simulation $\mathcal{S}(z)$ is the result of integrating \SFRD of each TNG over metallicity, while the analytical fits are directly modeled using the best-fit parameters for $\mathcal{S}(z)$ in Equation~\eqref{eq:sfr}.  

The choice of TNG simulation primarily affects the normalization and peak location of $\mathcal{S}(z)$. 
TNG50-1, with the highest mass resolution and smallest box size, yields the largest \ac{SFR} at all redshifts, while TNG300-1, the lowest resolution/largest box size run, is suppressed by roughly a factor $\sim 2$. The change of normalization in the \SFRD with numerical resolution is a well-known effect (see Appendix~A of \citealt{Pillepich2018b}). The effect is a result of higher-resolution simulations resolving more star formation by resolving more low-mass galaxies and star-forming gas cells and star particles in them.
This difference is encoded in $\mathcal{S}(z)$ normalization parameter $a$, which decreases with resolution (Table~\ref{tab:sfrd params}). 
Beyond the expected normalization change, the choice of TNG simulation also affects the redshift of peak star formation, where TNG50-1 peaks at $z \approx 2$, while TNG100-1 and TNG300-1 both peak near $z \approx 2.5$.  
The peak position is reflected in the parameter $c$, which emphasizes that the peak location of TNG50-1 differs from the other two simulations, which have approximately the same value of $c$. This may be an effect of the small box size and presence of a large cluster in TNG50-1, which might make TNG50-1 no longer representative of the ``average'' universe \citep{Pillepich2019}.
The parameters $b$ and $d$ control the low- ($z \lesssim 2$) and high-redshift ($z \gtrsim 2$) slopes of $\mathcal{S}(z)$, respectively. 
The low-redshift slope is fairly robust, but the analytical fit systematically overestimates $\mathcal{S}(z)$ at $z \lesssim 0.5$ by up to a factor of $\sim 1.5$.
In contrast, the high-redshift slope varies strongly with TNG simulation and shows a turning point near $z\approx 6-8$ that the analytical fit cannot reproduce (Figure~\ref{fig:sfr_z}). The fit over(under)estimates the high-redshift ($z\approx 6-12$) $\mathcal{S}(z)$ by up to a factor of $\sim 1.5$ for TNG50-1 (TNG100-1 and TNG300-1). At $z \approx 14$, the fit overestimates $\mathcal{S}(z)$ for all TNG simulations, by a factor of $\sim10$ for TNG300-1 and $\sim 2$ for TNG50-1 and TNG100-1. These effects may be a consequence of the optimization method, which prioritizes regions where the star formation rate is high (near the peak of star formation), and places less of an emphasis on regions in the parameter space where the star formation rate is low (e.g., at high redshift; see Sec~\ref{sec:optimization}).

We find that none of the TNG simulations align with the shape of the \citet{MadauFragos2017} $\mathcal{S}(z)$ across all redshifts. Among the TNG runs, TNG100-1 aligns best with the \citet{MadauFragos2017} $\mathcal{S}(z)$, consistent with the fact that the TNG simulations were calibrated at this resolution. However, the analytic form still fails to capture the steepening of the high-$z$ tail ($z \gtrsim 6$) for all simulations.

Turning to metallicity, Figure~\ref{fig:sfr_Z} shows $\mathcal{S}(Z)$, obtained by integrating \SFRD over redshift. 
We find that the analytical fits fail to represent the full simulation $\mathcal{S}(Z)$ distributions. 
Agreement is best around $10^{-1} \lesssim Z/Z_\odot \lesssim 10^{0}$ but both the peak and the tails are poorly represented by the assumed skewed log-normal form in the analytical fit. 
The simulation-based $\mathcal{S}(Z)$ has one peak near $Z/Z_\odot \sim 1$ and an additional bump feature near $Z/Z_\odot \approx 2-5$ that the fit is unable to capture for all TNGs. The analytical fit instead locates a single peak at $Z/Z_\odot \approx 1-2$, increasing with resolution. Following the second peak in the simulation $\mathcal{S}(Z)$, there is a sharp drop-off. The fit overestimates $\mathcal{S}(Z)$ at $Z/Z_\odot \gtrsim 6$ by up to a factor of $\sim10-1000$.
At the low-metallicity end, TNG50-1 shows a change in slope between $10^{-3} \lesssim Z/Z_\odot \lesssim 10^{-2}$ \citep[where $Z_\odot = 0.014$][]{Asplund2021}, while TNG100-1 and TNG300-1 show a flattening for $Z/Z_\odot < 10^{-2}$. This indicates that the fit to the metallicity distribution may be better at higher mass-resolution, though the dependence appears to be more complex considering the changes in slope of its low-metallicity tail.
A more accurate analytical fit representation would require a steeper high-metallicity cutoff and a low-metallicity tail with at least one additional change in slope. We emphasize that the fit to \SFRD is performed in two dimensions, and the optimization may affect the quality of fit in any given dimension alone, as compared to fitting a skewed log-normal distribution to the metallicity distribution alone (see Section~\ref{sec:optimization}).

The full two-dimensional structure of TNG100-1 \SFRD is illustrated in  Figure~\ref{fig:SFRD_Z_z_tng100}. We focus on TNG100-1 here, since it has intermediate resolution and box size and was the calibration resolution of the TNG suite (Section~\ref{sec:cosmological sim}). 
We find that the behavior of the fit is poor at both high metallicities ($Z\gtrsim \Zsun$) and at low metallicities and high redshift ($Z/Z_{\odot} \lesssim 10^{-2}$ and $z\gtrsim 8$).
At super-solar metallicities, the analytical fit diverges increasingly from the simulation at low redshifts due to a secondary bump in the simulation \SFRD that is smoothed over in the fit.
This leads to alternating over- and underestimates of $\mathcal{S}(z)$ on the high metallicity end of the peak (see the right panel of Figure~\ref{fig:SFRD_Z_z_tng100} above $Z/Z_\odot \gtrsim 1$). At the opposite end (low metallicity and high redshift), the analytical fit underestimates \SFRD compared to simulation, missing contributions at $Z/Z_{\odot} \lesssim 10^{-2}$ and $z\gtrsim 8$, as shown by the contours and in Figure~\ref{fig:sfr_Z}, which is analogous to the right panel but using a logarithmic scale for $\mathcal{S}(Z)$ (see Appendix~\ref{sec:3panelplots}. Note that there is a factor of $\approx0.6$ difference between the normalization of $\mathcal{S}(Z)$ plotted here as compared to Figure~\ref{fig:sfr_Z}. This is likely a result of the interpolation of \SFRD in log-space (see Section~\ref{sec:model}) and its effects remain to be explored. The qualitative behavior of the fit remains the same.
While this regime is not critical for the \ac{BBH} populations studied in Section~\ref{sec:BBH masses Z z}, it may be important for studies focused on very low-metallicity star formation at early times.

This shows that the discrepancy between the fit and the and the underlying \SFRD is greatest at high metallicities across redshift, and at low metallicities and high redshift. We attribute this to the differences in the form of the metallicity distributions of the fit and the underlying \SFRD of the TNG simulations (Figure~\ref{fig:sfr_Z}).

\subsection{Fit vs. simulation: Impact on the expected BBH merger rates}
\label{sec:BBH rates}

\begin{table}
    \centering
    \caption{Local merger rate at $z\approx0$.}
    \begin{tabular}{ccc}
    \hline
       TNG version  & $\mathcal{R}_\mathrm{sim}(z)$  & $\mathcal{R}_\mathrm{fit}(z)$\\
       &  [Gpc$^3$ yr$^{-1}$]  & [Gpc$^3$ yr$^{-1}$] \\
    \hline
        50-1 & 58.92 & 73.72 \\
        100-1 & 42.91  & 45.53 \\
        300-1 & 29.34  & 27.81 \\
    \hline
    \end{tabular}
    \label{tab:local-bbh-rates}
\end{table}

\begin{figure*}
\centering
\includegraphics[width=0.7\textwidth,angle=0]{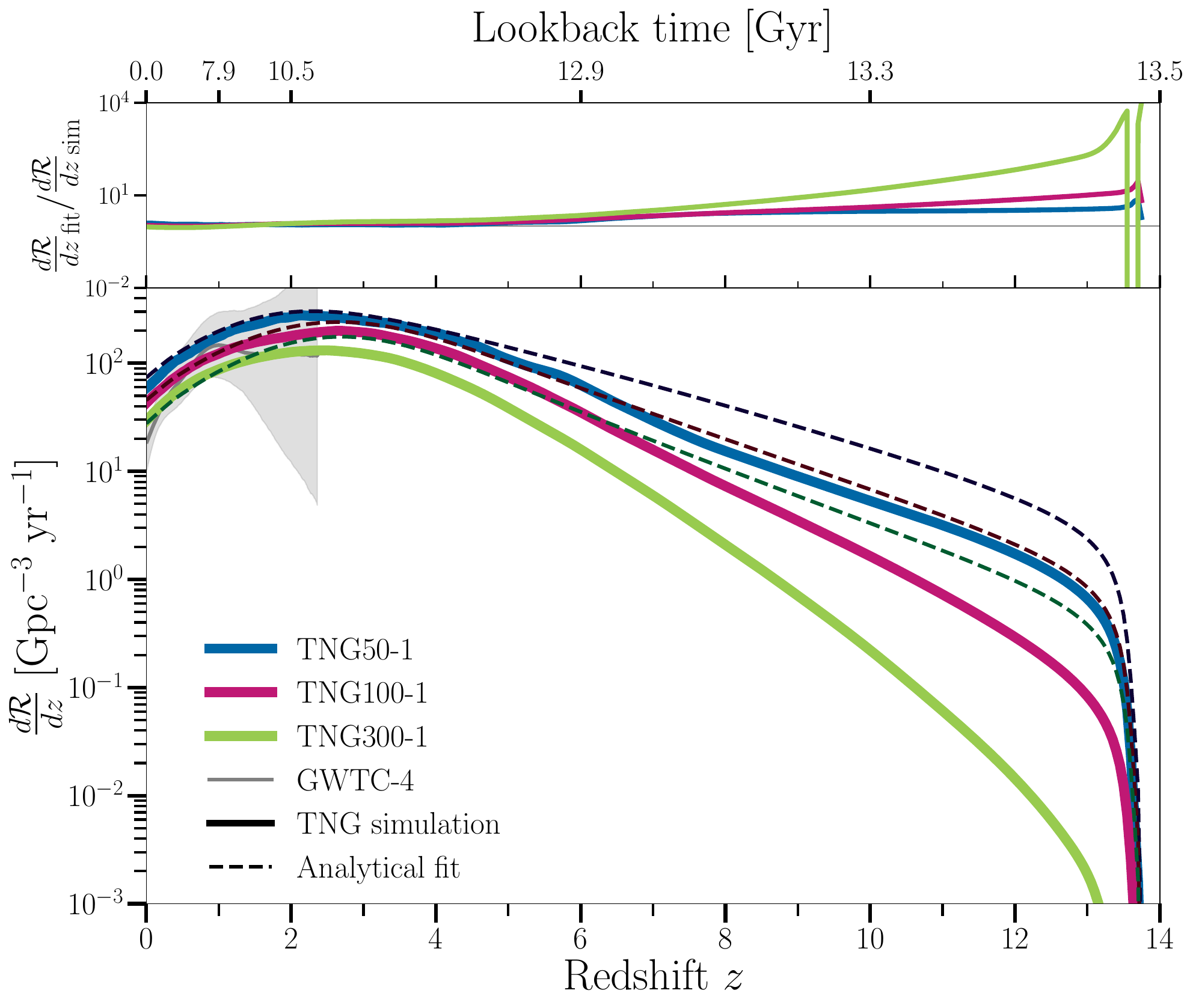}
\caption{\textbf{Bottom panel:} \ac{BBH} merger-rate density as a function of redshift. 
The observed \ac{BBH} merger rate based on GWTC-4 data from \citet{LVK2025} is shown in gray with the 95\% credible regions. 
\textbf{Top panel:}  the ratio between the rate from the analytical fit and TNG model (see Section~\ref{sec:comparison methods}). 
The noisy features in the ratio around $z\gtrsim13$ arise from division by very small numbers.
Colors and line styles match those in Figure \ref{fig:sfr_z}. 
}
\label{fig:merger rates}
\end{figure*}

We now investigate how \SFRD obtained from different TNG simulations and analytical fits (Figure~\ref{fig:sfr_z}) influences the modeled \ac{BBH} merger rates (calculated as in Section~\ref{sec:compas}). 
Figure~\ref{fig:merger rates} shows the resulting \ac{BBH} merger rate densities as a function of redshift, and Table~\ref{tab:local-bbh-rates} lists the corresponding local rates  at $z\approx0$.  
We find that the choice of TNG simulation significantly affects both the amplitude and the redshift evolution of the expected \ac{BBH} merger rate. 
At $z\approx0$, the local merger rate varies from $\sim 29$ to $\sim 59$  ${\rm Gpc}^{-3}{\rm yr}^{-1}$ between TNG300-1 and TNG50-1 when using the simulation \SFRD, and from $\sim28$ to $\sim74$ when using the analytical fit.
The merger rate peaks at $z \sim 2$ for TNG50-1 and at $z \sim 2.5$ for TNG100-1 and TNG300-1, similarly to the behavior of \SFRD (see Section~\ref{sec:SFRD results}).
Near the peak, the merger rate in TNG50-1 exceeds that in TNG300-1 by roughly a factor of three (Figure~\ref{fig:merger rates}). 
At high redshift ($z \gtrsim 6$), the TNG50-1 merger rate is up to two orders of magnitude larger than in TNG300-1.

When using the analytical fit, the TNG50-1 merger rate at $z\approx0$ is overestimated by about $15 {\rm Gpc}^{-3}{\rm yr}^{-1}$. 
For TNG100-1 and TNG300-1, the local rates from the simulation \SFRD and the fit agree within $\pm 2 {\rm Gpc}^{-3}{\rm yr}^{-1}$ (Table~\ref{tab:local-bbh-rates}).
The fit reproduces the merger-rate evolution most accurately between $z \approx 1$--$4$ for TNG50-1, $z \lesssim 2$ for TNG100-1, and $z \lesssim 1.5$ for TNG300-1. This behavior broadly reflects the quality of the fit to the underlying \SFRD (see Section~\ref{sec:SFRD results}). However, the deviations between fit and simulation at $z \lesssim 1$ are largely suppressed in the merger rate predictions, and the fit does not systematically overestimate or underestimate the $z\lesssim 1$ merger rates, likely due to the mixture of binaries with different delay times that formed at a range of redshifts.
At high redshift ($z \gtrsim 6$), the analytical fits yield systematically larger \ac{BBH} merger rates than those derived directly from the TNG  simulation \SFRD---by factors of up to $10^4$ for TNG300-1 and $\sim10$ for TNG50-1 and TNG100-1 (Figure~\ref{fig:merger rates}). 
Using the fit also leads to a shallower slope in the $z\gtrsim 4$ regime and a steeper decline in the merger rate toward $z \sim 14$.

\subsection{Fit vs. simulation: Impact on BBH masses}
\label{sec:BBH masses}

\begin{figure*}[t]
    \centering
    \includegraphics[width=0.9\textwidth,angle=0]{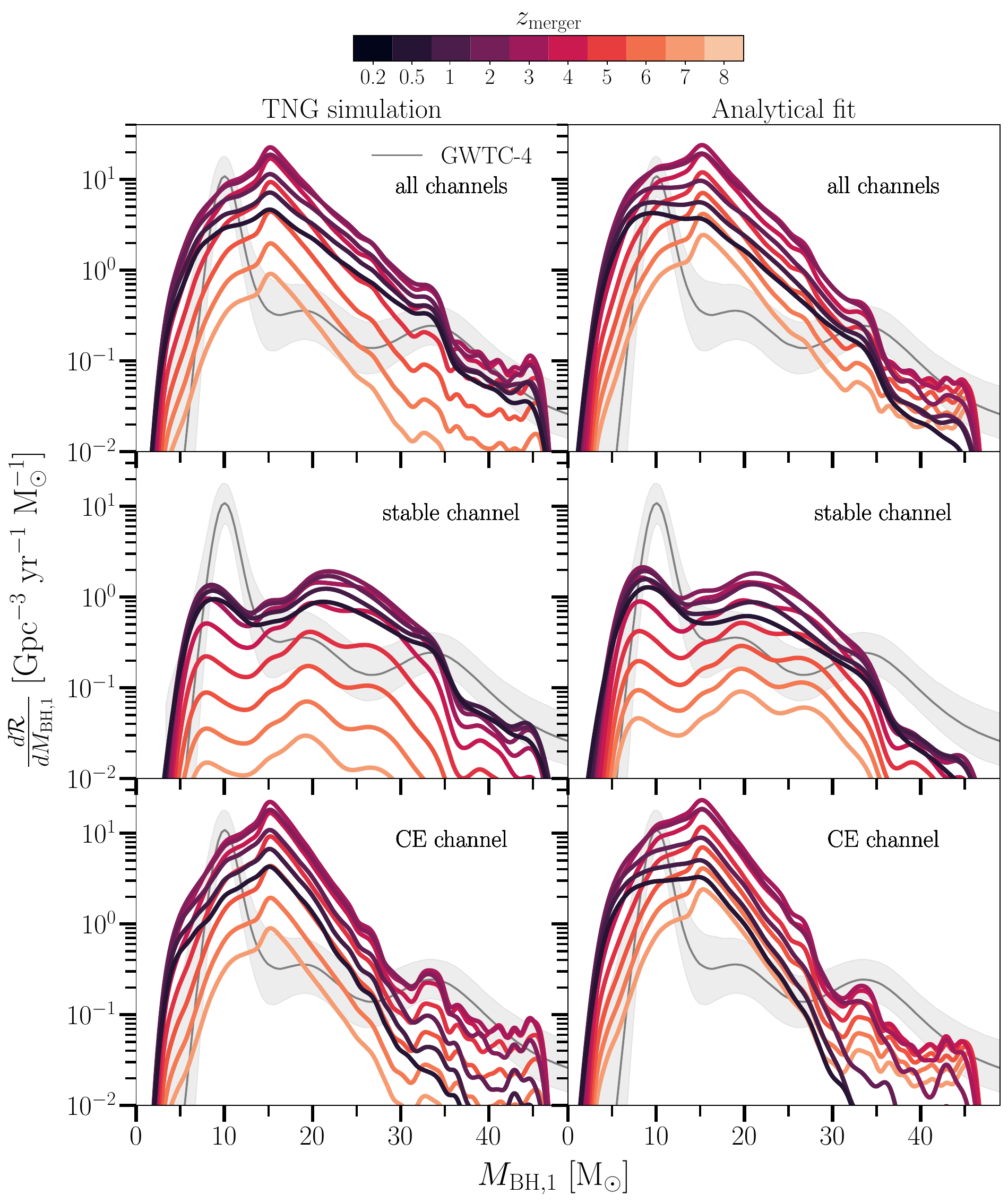}
    \caption{Redshift evolution of the \ac{BBH} primary mass distribution, $M_{\rm{BH,1}}$ [${\rm Gpc}^{-3}{\rm yr}^{-1}\Msun^{-1}$], for the TNG100-1, computed  using the full simulation \SFRD (left column) and the analytical fit (right column). The darkest color represents the local mass distribution $z_\mathrm{merger} = 0.2$ and the lightest is at $z_\mathrm{merger}=8$.
    The B-spline \ac{BBH} primary mass distribution from GWTC-4 \citep{LVK2025} is shown in gray. 
    The rows represent different formation channels: the top row includes mergers from both the stable mass transfer and common-envelope (CE) channels, the middle row shows only stable mass transfer channel mergers, and the bottom row includes only CE channel mergers.}
    \label{fig:formation channels}
\end{figure*}

\begin{figure*}[t]
\centering
    \includegraphics[width=0.9\textwidth,angle=0]{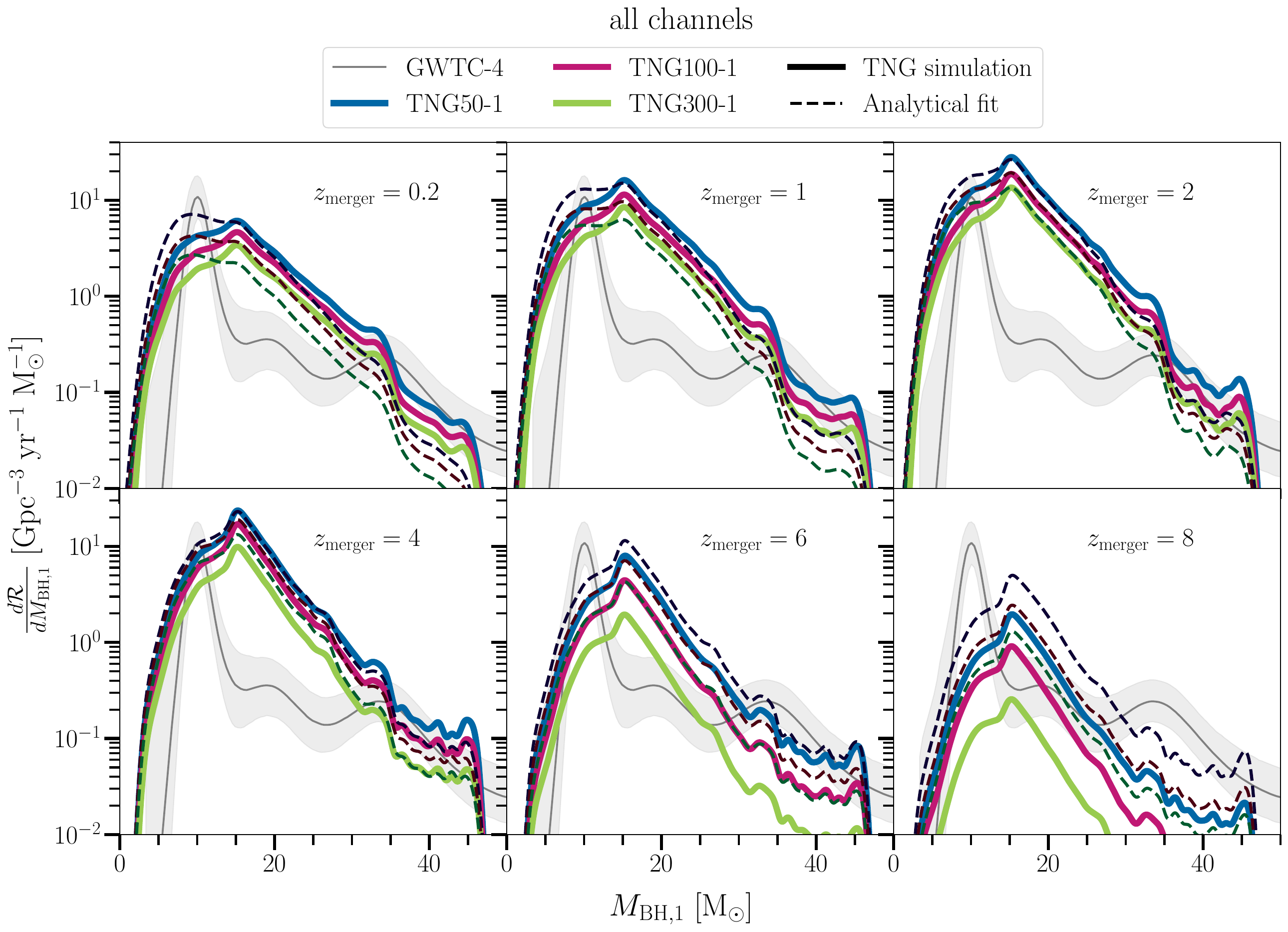}
    \caption{Comparison of \ac{BBH} primary mass distributions (in units of ${\rm Gpc}^{-3}{\rm yr}^{-1}\Msun^{-1}$) for the TNG simulations. The colors and line styles are the same as in Figure \ref{fig:sfr_z}. 
    For reference, the observed local \ac{BBH} primary mass distribution at $z_\mathrm{merger} = 0.2$ from GWTC-4 \citep{LVK2025} is shown in gray. 
    The panels show the mass distributions for redshifts $z_\mathrm{merger}=0.2,1,2,4,6,8$.}
        \label{fig:massdist comparison}
\end{figure*}

We now investigate the impact of using the analytical fit versus simulation-based \SFRD on the primary mass distribution of \acp{BBH} in Figures~\ref{fig:formation channels} and \ref{fig:massdist comparison}. 
Across redshift, the mass distribution typically shows several key features: a steep incline from $\sim 5\,M_\odot$ to a first local peak feature around $8\,M_\odot$, followed by a further incline to a peak around $16\,M_\odot$, a power law-like decline with an approximately consistent negative index between $16\,M_\odot$ and $33\,M_\odot$, and a further decline with an additional final local peak/mass feature around $45\,M_\odot$ (Figure~\ref{fig:formation channels}).

These features are formed by the varying contributions from the two formation channels: stable mass transfer and CE (Section~\ref{sec:compas}). 
The $8\,M_\odot$ feature is primarily attributed to the stable channel, while the $16\,M_\odot$ peak and $\sim 33$ and $45\,M_\odot$ features are dominated by the CE channel \acp{BBH} (Figure~\ref{fig:formation channels}). 
The evolution across redshift is relatively self-similar with the rate increasing toward the peak of star formation near $z \approx 2$, followed by a decline towards higher redshift. 
The CE mass distribution at $\gtrsim 30\,M_\odot$ increases with redshift, though both channels contribute to the high-mass features (Figure~\ref{fig:formation channels}). They are a result of pulsational-pair instability effects, and therefore appear in both channels \citep[see Figure 6 of][]{vanSon2022a}. The $33 \Msun$ feature is an artificial feature produced by the switch between core-collapse supernova and pair pulsation prescriptions. The feature at $45 \Msun$ is the pileup from pair-instability supernova. At higher redshifts ($z \gtrsim 2 $), stochastic features appear for  $M_\mathrm{BH, 1} \gtrsim 30\,M_\odot$ due to sampling. 
All mass features are present across redshift, although their relative prominence varies due to the changing formation channel contributions (Figure~\ref{fig:formation channels}). 

Using the analytical \SFRD affects the \ac{BBH} mass distribution. 
We find that the analytical mass distribution experiences less evolution in the overall rate for the total distribution and both channels (consistent with Figure~\ref{fig:merger rates}).
The analytical \SFRD also affects both channels. In the stable channel, the overall normalization of the mass distribution is overestimated, most prominently at higher redshift (Figure~\ref{fig:formation channels}).
In the CE channel, the $8\,M_\odot$ feature is enhanced due to the increase of the shoulder in the CE channel mass distribution at low masses. This causes the low-mass peak to broaden into a plateau at $z \approx 0.2$.


This excess persists across TNG simulation.
The \ac{BBH} mass distribution produced by the analytical fit shows significant deviations compared to the TNG simulation \SFRD that the fit is trying to represent (Figure~\ref{fig:massdist comparison}). 
Most notably, at $z \lesssim 2$, the fit results in an artificial bump feature at $M_\mathrm{BH, 1} \sim 8\,M_\odot$ and suppresses both the high mass peaks for all TNG simulations (Figure~\ref{fig:massdist comparison}). 
At $\sim8\,M_\odot$, the fit results in a factor of $\sim 2-3$ greater \ac{BBH} $d\mathcal{R}/dM$ for all TNG simulations. At high masses, the rate is a factor of $\sim 3-5$ lower at $M_\mathrm{BH, 1} \gtrsim 35$ (Figure~\ref{fig:massdist comparison}; also see Appendix~\ref{sec:difference}). 
The locations of the other mass features are more consistent across fit and simulation \SFRD.

Consistent with the results of Section~\ref{sec:BBH rates}, the rate decreases with simulation resolution, and there is no variation in the locations of mass features across TNG simulation.
The analytical fit increasingly overestimates the rate across all masses as redshift increases, by up to a factor of $\sim 3,  4, 10$ at $z=8$ for TNG50-1, TNG100-1, and TNG300-1, respectively (see Appendix~\ref{sec:difference}).

We note that the mass distributions produced using our models do not reproduce the primary mass distribution as observed by \citet{LVK2025}, regardless of \SFRD modeling assumptions. 
The peak at $16 \Msun$ across redshift (Figure~\ref{fig:massdist comparison}) does not coincide with any features in the observed mass distribution. This peak is the global maximum across redshift when using the TNG simulation \SFRD to model the \ac{BBH} population and is attributed to the CE channel (Figure~\ref{fig:formation channels}), appearing for both mass distributions modeled using TNG and fits. 
The analytical fit, which results in a more prominent $8 \Msun$ feature that creates a plateau between the $8-16 \Msun$, nevertheless underpredicts the low-mass peak by a factor of 1.5-4. Only the mass distribution modeled using the fit to TNG50-1 is consistent with the LVK rate at $8 \Msun$, within the 95\% credible region. 
The form of the stable mass transfer channel more closely resembles the observed distribution, including the a low-mass peak at $\sim8 \Msun$, a feature at $\sim20 \Msun$, and extending towards higher mass. However, its $8 \Msun$ is approximately a factor of 10 too low, and the $20 \Msun$ feature is approximately a factor of 1.5 too high.

This is an expected result of the binary evolution model used in this work. See \citet{vanSon2022a} and \citet{vanSon2022b} for discussion regarding the contribution of the formation channels to the local \ac{BBH} merger population using the same binary evolution model as used here.
The locations of key features in the primary mass distribution modeled using a given binary evolution model are robust across \SFRD modeling variations, consistent with \citet{vanSon2023}. The location of these features also do not show significant redshift evolution \citep{vanSon2022a}, though the relative contributions of the CE and stable mass-transfer channels varies across redshift.
Our results agree with these previous findings. This shows that binary modeling assumptions have a stronger impact on the locations of features in the \ac{BBH} mass distribution, while \SFRD can impact the overall rate (normalization) or their relative prominence (Figure~\ref{fig:massdist comparison}). 
This emphasizes that changes in binary evolution modeling are likely required in addition to varying \SFRD in order to obtain a mass distribution more consistent with the observed distribution. See Section~\ref{sec:obscomparison} for further discussion of how other population synthesis results compare to LVK observations.

\subsubsection{Mass distribution as a function of formation metallicity and redshift}
\label{sec:BBH masses Z z}

\begin{figure*}
\centering
    \includegraphics[width=0.8\linewidth,angle=0]{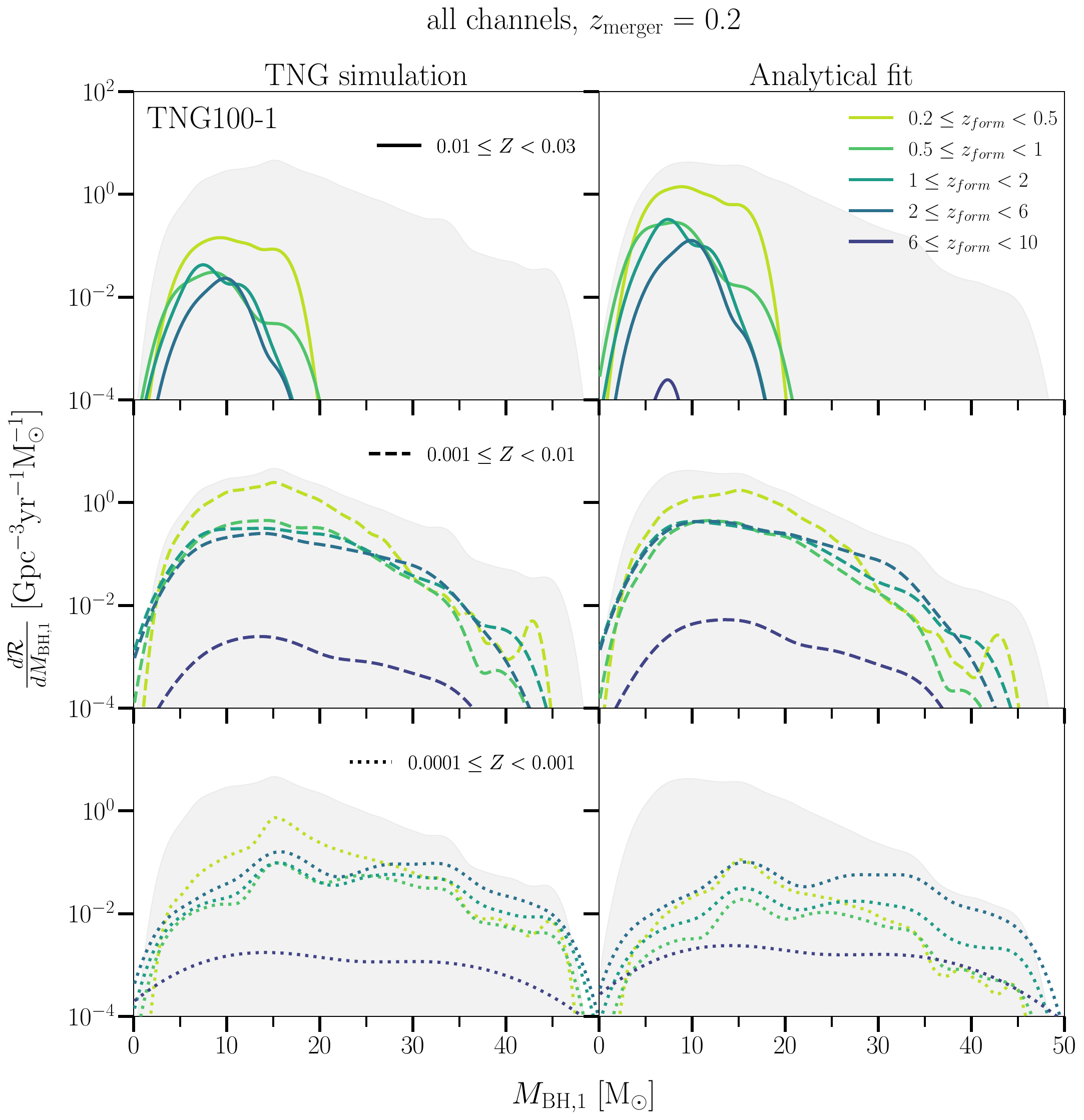}
    \caption{The $z_\mathrm{merger} = 0.2$ \ac{BBH} primary mass distribution for TNG100-1 using \SFRD from the simulation (left) and the analytical fit (right), binned by formation redshift (color) and metallicity (line style). \textbf{Top row:} the solid lines represent the highest metallicity bin ($0.01 \leq Z < 0.03$); \textbf{middle row:} the dashed lines represent the intermediate metallicity $0.001 \leq Z < 0.01$ bin; \textbf{bottom row:} the dotted lines are the lowest metallicity bin ($0.0001 \leq Z < 0.001$). The gray shading represents the total rate, and the behavior at the edges is attributed to the use of KDEs in plotting the mass distribution. The bins cover the full parameter space of binaries in our population that merge at $z_\mathrm{merger}=0.2$.}
    \label{fig:massdist_Z_zform}
\end{figure*}

\begin{figure}
    \centering
    \includegraphics[width=1\linewidth,angle=0]{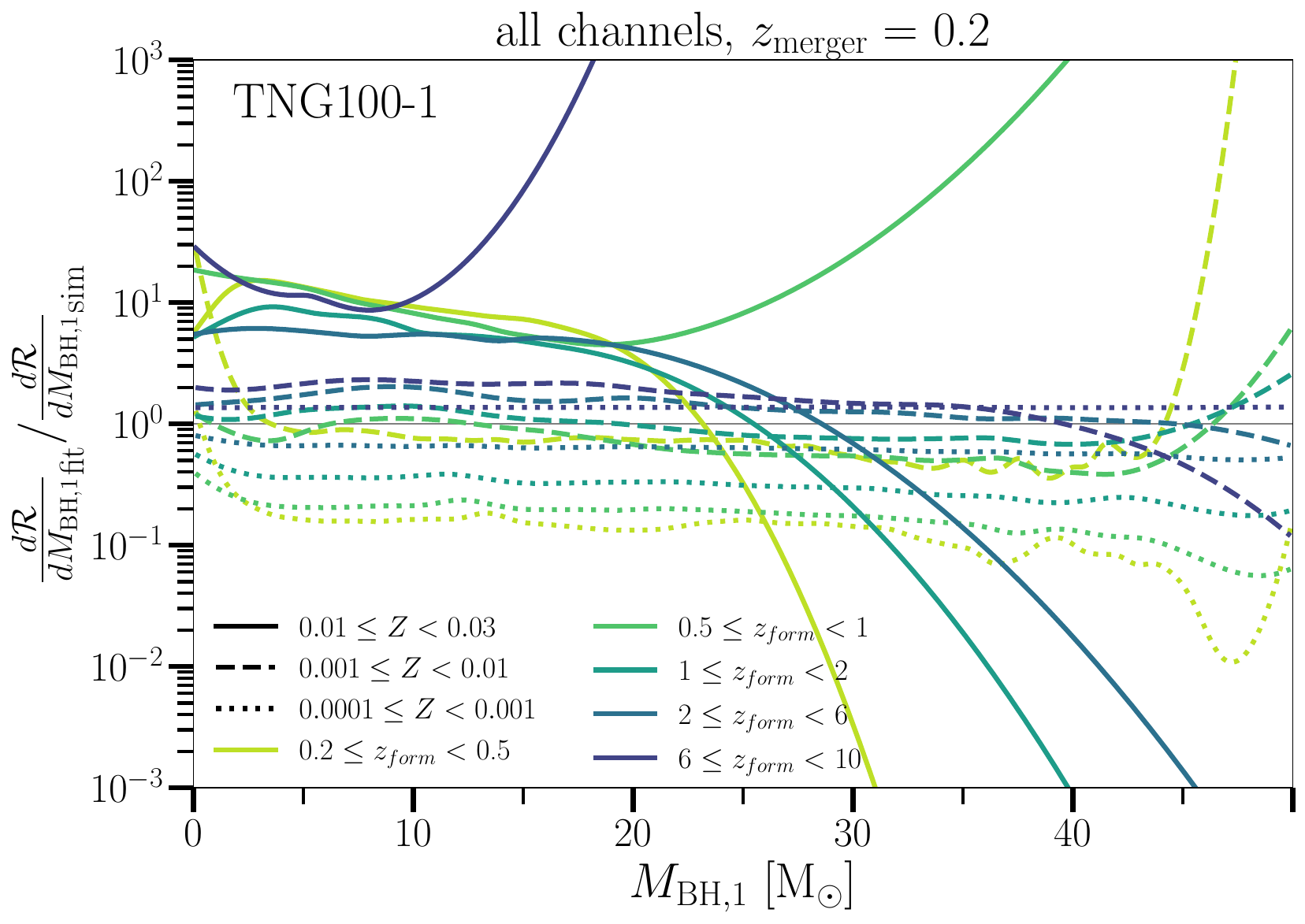}
    \caption{The ratio between the fit and simulation mass distributions, binned by metallicity and formation redshift. The colors and line styles are analogous to Figure~\ref{fig:massdist_Z_zform}. The behavior of the low metallicity (solid) distribution ratios at $M_\text{BH, 1} \gtrsim 15$ are due to their very small rates in that regime.}
    \label{fig:massdist_Z_zform_error}
\end{figure}

To investigate the origin of the artificial bump in the primary mass distributions around $8 \Msun$ obtained using the analytical \SFRD fit (Figures~\ref{fig:formation channels} and \ref{fig:massdist comparison}), we examine the formation redshifts and metallicities of the binaries contributing to this feature for TNG100-1. We bin the $z_\mathrm{merger}=0.2$ primary mass distribution for both the analytical fit and the TNG \SFRD to discern whether binaries forming in a specific region of \SFRD contribute most to the excess rate near $8 \Msun$.
Figure~\ref{fig:massdist_Z_zform} illustrates the contributions of different metallicity and formation redshift bins to the \acp{BBH} mass distribution, with the corresponding ratios between the fit and simulation mass distributions for each bin in Figure~\ref{fig:massdist_Z_zform_error}.

We find that the analytical fit produces an excess in the highest metallicity bin ($0.01 \leq Z \leq 0.03$) at $M_\mathrm{BH, 1} \lesssim 15$, across all formation redshifts, compared to models using the full simulation \SFRD (Figure~\ref{fig:massdist_Z_zform}). 
The peak of the primary BH mass distribution in this metallicity range coincides with the artificial feature seen in the primary mass at $z_\mathrm{merger} \lesssim 2$. 
In contrast, the lowest metallicity ($10^{-4} \leq Z \leq 10^{-3}$) mass distributions are underestimated by the analytical fit by up to a factor $\sim 10$ at low formation redshift, but the difference between their contributions systematically decreases with increasing redshift (Figure~\ref{fig:massdist_Z_zform_error}). 
The intermediate metallicity bin ($10^{-3} \leq Z \leq 0.01$) is least affected, showing a flat ratio between fit and simulation with at most a factor of $\sim 2$ across all redshifts. 

These trends indicate that the analytical fit overproduces the rate of high-metallicity ($Z \gtrsim Z_\odot$) low-mass \acp{BBH} and underpredicts the contribution at low metallicities. This is consistent with the poor performance of the analytical fit at super-solar metallicities as seen in Figure~\ref{fig:SFRD_Z_z_tng100} (we show the formation redshift and metallicity bins overlaid onto \SFRD of TNG100-1 in Figure~\ref{fig:SFRD_Z_z_tng100_regions} in Appendix~\ref{sec:2dplotregions}).
The highest metallicity bin modeled using our \texttt{COMPAS} models contains the peak of \SFRD. This coincides with the metallicities where the analytical fit fails to reproduce the metallicity distribution due to a secondary, high-metallicity feature (Section~\ref{sec:SFRD results}). While \texttt{COMPAS} cannot produce binaries with $Z > 0.03$, this secondary bump causes the analytical fit to over-predict $\mathcal{S}(z)$ between the two features. This may result in the elevated high metallicity rate in the primary mass distribution (Figure~\ref{fig:massdist_Z_zform}).

At high redshift and low metallicity, the analytical fit underestimates the fit to \SFRD, leading to lower \ac{BBH} merger rates and suppressed peaks in the high-mass end of the mass distribution (Figure~\ref{fig:massdist_Z_zform}). 
Consequently, the contribution from the lowest metallicity bin drops by up to a factor $\sim 2$. 
As shown in Figure~\ref{fig:massdist_Z_zform}, these low-metallicity populations span the full range of primary BH masses, but the overprediction of higher metallicity systems at $M_\mathrm{BH, 1} \lesssim 20$ makes the discrepancy most prominent at higher masses.
The intermediate metallicity bin $10^{-3} \leq Z \leq 10^{-2}$ (dashed lines) remains relatively unaffected across most redshifts, though it shows an increased contribution (by about a factor 2) at the highest redshifts  ($6 \lesssim z  \lesssim 10$).

We therefore caution that models relying on analytical \SFRD fits may be unreliable, especially when considering contributions from populations originating at high redshift ($z \gtrsim 6$) and low metallicity ($Z/Z_\odot \lesssim 10^{-2}$) (Figure~\ref{fig:massdist_Z_zform}). 
Improving the fit to \SFRD in both the high metallicity-low redshift and the low metallicity-high redshift regime is essential, since a substantial fraction of binaries merging at $z_\mathrm{merger}=0.2$ form in these regions. 
Future analytical fits should allow for more flexible modeling of the metallicity distribution---potentially incorporating a secondary high-metallicity peak and a flatter low-metallicity tail---to better capture the structure of \SFRD.

\section{Discussion}
\label{sec:discussion}
\subsection{Comparison to earlier work investigating the impact from \SFRD on BBH populations}
\label{sec:comparisontoearlierwork}
\subsubsection{Analytical fit \SFRD}
Analytical fits are a convenient means of representing cosmological simulation \SFRD distributions due to their computational simplicity and the ability to quantify \SFRD uncertainties using physically-motivated parameters. 
Earlier work has explored the impact from using analytical fits for \SFRD assumptions on double compact object merger populations. 
\citet{Neijssel2019} explored the impact of \SFRD using analytical fits, using a parametrized prescription that combines the \citet{MadauDickinson2014} functional form for $\mathcal{S}(z)$ with a log-normal metallicity distribution derived from the \ac{GSMF} and \ac{MZR}. 
The authors obtain a ``preferred model'' by best-fitting analytical fit parameters from \ac{BBH} observations from the first two observing runs of \ac{LVK}. They compare this against models for $\mathcal{S}(z)$, \ac{GSMF}, and \ac{MZR} from galaxy observations \citep{Panter2004, Savaglio2005, LangerNorman2006} or cosmological simulations \citep{Furlong2015, Ma2016}. 
They find a \ac{BBH} rate for the preferred model that is about an order of magnitude lower than those produced using the other (analytical, but observation-based) models, which is attributed to the ``preferred'' model having larger metallicities at low redshift, which decreases the \ac{BBH} yield at low redshift. The \citet{Neijssel2019} analytical form and fit parameters are a widely used \SFRD model \citep[e.g.,][]{COMPAS, Stevenson2019, Riley2021, Broekgaarden2022, vanSon2022a, vanSon2023, Romero-Shaw2023, Roy2025}. 
However, \citet{Chruslinska2024} showed that the preferred model of \citet{Neijssel2019} is unphysical due to the negligible low-metallicity star formation at $z \lesssim 2$, and the resulting \ac{BBH} rates deviate significantly from the rates produced using observational and cosmological simulation \SFRD assumptions.  

\citet{Briel2022} model  metallicity-dependent transient rates (including \ac{BBH} mergers) using the TNG100-1, EAGLE \citep{Schaye2015}, and MilliMillennium \citep{Springel2005} cosmological simulations. 
They compare the full \SFRD from each simulation to an empirical analytical model combining the \citet{MadauDickinson2014} $\mathcal{S}(z)$ with the \citet{LangerNorman2006} metallicity distribution. 
While the simulations broadly agree at $z\lesssim 2$, they diverge at higher redshifts. 
Consistent with our findings, none of the simulation \SFRD are well-represented by the empirical model; in particular, the high-redshift  slope deviates substantially from a \citet{MadauDickinson2014} form. Moreover, the \citet{LangerNorman2006} metallicity prescription does not reproduce the metallicity distributions found in cosmological simulations. 
%
The resulting \ac{BBH} merger rates in \citet{Briel2022} vary by nearly  an order of magnitude, with the peak location spanning between $1.5 \lesssim z \lesssim 3$, depending on the cosmological simulation used. They conclude that metallicity-dependent transients with long delay times, such as \ac{BBH}, are most sensitive to \SFRD assumptions,
attributed primarily to differences in the metallicity distribution and its redshift evolution, which are not captured by the \citet{LangerNorman2006} model. 
Their \ac{BBH} merger rate based on TNG100-1 is consistent with our results using the full TNG100-1 \SFRD, including at $z \approx 0$, despite differences in  population synthesis model (\texttt{BPASS} versus \texttt{COMPAS}). 

\citet{vanSon2023} create an analytical fit for \SFRD model based on the TNG100-1 simulation---this is the work that our paper builds forward on.
In this paper the authors focus on investigating the performance of the analytical fit as a proxy for the TNG100-1 simulation \SFRD by comparing the accuracy for \SFRD (but not the \ac{BBH} populations.) The work builds on the approach from \citet{Neijssel2019}, but allows for a skewed log-normal metallicity distribution with a low-metallicity tail in $\text{log}_{10}Z$.
Their analytical fit is able to capture the large-scale behavior of the TNG100-1 \SFRD, but similarly fails to fit a secondary bump in $\mathcal{S}(Z)$ (see the bottom left panel of their Figure 1). 
The authors additionally investigate how changing the analytical fit parameters impacts the \ac{BBH} population. 
They find that there is little variation in the locations of major features in the \ac{BBH} primary mass distribution over variations in \SFRD parameters. Similarly, we find that locations of major features are consistent across TNG version and redshift (see Figures~\ref{fig:formation channels}-\ref{fig:massdist comparison}).
\citet{vanSon2023} find that the low-mass end of the \ac{BBH} mass distribution is least affected by \SFRD, while the high-mass end is most affected by parameters that describe the redshift evolution of \SFRD. We find that both the low-mass peak and high-mass end of the mass distribution is affected by the redshift evolution of \SFRD (see Figure~\ref{fig:massdist comparison}).

Their $z=0.2$ merger rates for the stable and CE channels are 17.1 and 41.8 mergers Gpc$^{-3}$ yr$^{-1}$, respectively, when using the fiducial \SFRD. Depending on \SFRD parameter variations, the merger rates range between $6.3-65.4$ Gpc$^{-3}$ yr$^{-1}$ for the stable channel and $24.2-144.1$ Gpc$^{-3}$ yr$^{-1}$ for the CE channel.
Our merger rates for all three TNG versions are approximately within the total range found by \citet{vanSon2023}, though on the lower end, as we calculate the merger rate at $z\approx 0$ rather than $z= 0.2$. For the fitted TNG100-1 \SFRD, we find a total rate of 45.53 Gpc$^{-3}$ yr$^{-1}$ at $z\approx 0$ (see Table~\ref{tab:local-bbh-rates}).
%
Minor differences in fit values likely arise from differing cosmological parameters (Planck 2016 to match TNG, while \citet{vanSon2023} use \citealt{Planck2018}) and differences in assumed onset redshift of star formation ($z=14$ in our {\tt COMPAS} setup versus $z=10$ in \citealt{vanSon2023}).

A similar analysis to this work should be extended to other cosmological simulations in order to verify whether other \SFRD distributions may be more reasonably represented by analytical prescriptions than TNG. If this is the case, improvements of the analytical fit could be explored. This may include features such as allowing for a secondary, redshift-dependent peak in the metallicity distribution and greater flexibility for changes in slope within the low-metallicity tail. Improvements in optimization methods should also be explored, as they must balance between an accurate fit to the peak of star formation and the low-metallicity tail of the metallicity distribution. 

\subsubsection{Simulation-based \SFRD}

Cosmological simulations have also been directly (i.e., without using a fit) used in population synthesis studies of \ac{GW} sources. This method avoids constructing an explicit \SFRD: \acp{BBH} formed in population synthesis are associated with newly formed star particles of similar metallicity, and their merger time is derived from the formation redshift and delay time.
For example, \citet{Mapelli2017, Mapelli2019}
calculate \ac{BBH} merger rates via Monte Carlo placement of \ac{BBH} systems simulated using a population synthesis model using \texttt{MOBSE} \citep{Giacobbo2018a}, an updated version of the \texttt{BSE} code \citep{Hurley2000, Hurley2002}, onto star particles in cosmological simulations.  
Using the Illustris-1 simulation \citep{Vogelsberger2014}, \citet{Mapelli2017} find local \ac{BBH} merger rates typically an order of magnitude higher than ours, attributable to differences in binary physics rather than \SFRD assumptions. 
\citet{Mapelli2019} find minimal redshift evolution in the mass distribution across several binary evolution variations, which is attributed to \acp{BBH} from low-metallicity ($Z \lesssim 4 \times 10^{-3}$) dominating the \ac{BBH} population across redshift.

\citet{Artale2019, Artale2020} apply a similar method to the EAGLE simulations to explore the \ac{BBH} merger rate per galaxy and its dependence on galaxy mass, \ac{SFR}, and metallicity. They find that the stellar mass of a galaxy is the strongest tracer of the double compact object merger rate per galaxy. \citet{Artale2020} show that specific \ac{SFR} variations (where the specific \ac{SFR} is the \ac{SFR}/galaxy stellar mass) can cause order of magnitude variations in the \ac{BBH} merger rate. Our values for the local merger rate given the TNG \SFRD variations fall within that range. 

\citet{Marinacci2025} incorporate double compact object calculations directly into the moving-mesh code {\tt AREPO} (both on-the-fly and in post-processing). They demonstrate their approach using MillenniumTNG and explore how box size affects \SFRD and the resulting \ac{BBH} populations, finding minimal sensitivity to simulation volume. While on-the-fly modeling has conceptual advantages, the computational cost likely limits parameter exploration relative to post-processing methods. 

\subsubsection{Alternative \SFRD models}

The \ac{GW} source populations have also been modeled using observational galaxy scaling relations with {\tt GalaxyRate} \citep{Santoliquido2022} for \SFRD and {\tt SEVN} \citep{Spera2017, Spera2019}, including variations in the \ac{GSMF} \citep{Chruslinska2019b}, galaxy main sequence \citep{Speagle2014, Boogaard2018, Popesso2023}, and metallicity relations \citep{Mannucci2011, Andrews2013, Curti2020}.
These authors model \SFRD analytically following \citet{MadauFragos2017} and a redshift-dependent metallicity spread using {\tt CosmoRate} \citep{Santoliquido2020}.
The resulting \ac{BBH} merger rates vary by about an order of magnitude across model variations and generally exceed the observed rate \citep{LVK2025}. They find that binary evolution assumptions have stronger impact on \ac{BBH} rates than \SFRD variations, and that analytical \SFRD models can mimic more detailed galaxy-based ones when allowing a large metallicity spread. Their \ac{BBH} rates exceed ours (based on TNG and {\tt COMPAS}) by roughly an order of magnitude. 
Whether population synthesis differences alone can explain the rate offset, or whether cosmological simulations underpredict the \ac{BBH} rate relative to observation-based models, remains an open question. 

A direct comparison of our results with observational \SFRD distributions, such as those derived by \citet{Chruslinska2019b}, would identify whether the features where the analytical fit fails for TNG also exist in observations.
If those common features exist, this would emphasize the need to move away from representing star formation histories using \SFRD and constrain the behavior and parameter space that realistic \SFRD should span. 

\subsection{Effects from cosmological simulations and binary evolution modeling} 
\subsubsection{Resolution effects}

Most cosmological simulation suites include multiple runs with varying resolution and simulation box volume. In a given simulation, gas cell masses span a factor of $\sim2$ around the target mass resolution \citep{Pillepich2018b}. 
IllustrisTNG exhibits a $\sim30\%$ systematic shift in stellar mass for each factor-of-eight change in mass resolution at fixed halo mass (see Appendix~A of \citealt{Pillepich2018b}).
In this work, we found that numerical effects (e.g., mass resolution and simulation box volume) in the IllustrisTNG simulations affect their \SFRD and the resulting \ac{BBH} populations, but future studies should investigate this effect systematically and for other simulations. As both resolution and box size vary with each simulation, isolating the effects of resolution versus box size requires a larger set of cosmological simulations. We will perform an in-depth study of the effects of resolution and volume on \SFRD and the resulting \ac{BBH} merger population in future work.

In considering these numerical effects, it is important to note that higher resolution does not necessarily imply a more reliable star formation history. This is for several key reasons. First, cosmological simulations use subgrid modeling prescriptions for processes that are below the resolution of the simulation (Section~\ref{sec:cosmological sim}). This means that for the processes that are modeled using subgrid models, such as supernova feedback or the interaction of stellar winds with the interstellar medium, increasing the mass resolution indefinitely will not necessarily make its \SFRD more reliable as long as the same subgrid models continue are used. 
Additionally, increasing the resolution often comes at the cost of decreasing the box size due to computational feasibility. This increases the effect of cosmic variance, restricts the size of structures (e.g., galaxy clusters) that can form, and limits the statistics of the simulation, which means that rarer objects are less sampled. 
However, higher mass resolution allows more low-mass dwarf galaxies to become resolved and therefore increases the star formation rate, allowing star formation to start earlier. As both high redshifts and low metallicities are favorable to the formation of merging \acp{BBH}, it may therefore still be advantageous to use higher resolution cosmological simulations. 

Second, cosmological simulations are calibrated to match a subset of scaling relations. The choice of relations and the degree of allowed freedom varies across simulation suites. IllustrisTNG, for example, was calibrated to reproduce the redshift evolution of the star formation rate density, the $z\approx0$ \ac{GSMF}, the black hole mass–stellar mass relation, the galaxy size-mass relation, and the \ac{SMHM} relation \citep{Pillepich2018b}, with some flexibility. In contrast, EAGLE simulations \citep{Schaye2015} were calibrated primarily to reproduce the \ac{SMHM} relation with additional constraints from the \ac{GSMF}, galaxy size-mass relation, and black hole scaling relations.
As the TNG simulations were calibrated at the TNG100-1 resolution, the higher-resolution TNG50-1 produces excess star formation and therefore a higher \SFRD normalization compared to observations (Figure~\ref{fig:sfr_z}).

This reflects IllustrisTNG's use of strong resolution convergence: parameters are not recalibrated at higher resolution, preserving numerical and physical consistency across runs but not compensating for resolution-driven changes in subgrid physics \citep{Pillepich2018b}.
\citet{Marinacci2014} find that strong convergence across resolution leads to more consistent galaxy properties in some contexts. By contrast, the EAGLE simulations adopt weak resolution convergence, recalibrating subgrid parameters at higher resolution to minimize systematic shifts \citep{Schaye2015}.
Whether weak or strong convergence is preferable depends on which scaling relations were used for the (re)calibration, complicating whether higher resolution can or cannot be assumed to be more reliable. A systematic comparison of weak versus strong convergence effects on \SFRD and \ac{BBH} populations remains unexplored.

Evaluating the performance of the analytical \SFRD model as an approximation of the TNG simulations is the first step of investigating the effect of star formation histories on \ac{BBH} populations across different cosmological simulations, where physical assumptions also vary alongside numerical effects. It allows us to determine whether such fits can be a reliable means of comparing the complex, 2-D \SFRD of different cosmological simulations. Following our results, analytical fits may still be applicable to comparing global trends across simulations, but may miss features incompatible with the form of the model. Future work should evaluate whether other cosmological simulations, beyond IllustrisTNG, may be better represented using analytical forms such that is used in our analysis. Identifying common features across different cosmological simulations and observed star formation histories, and whether they align with analytical \SFRD models, can better inform which cosmological simulations are best suited to \SFRD modeling and whether it is necessary to update analytical models.


A more complete treatment must move beyond an averaged, cosmic \SFRD to include the effects of types of different galactic environments: galaxy characteristics such as mass and the spatial gradients and scatter in metallicity and \ac{SFR} within galaxies. The former would require associating star particles with their host subhaloes to determine galaxy association, while exploring the latter would likely require zoom-in simulations such as FIRE and FIREbox \citep{Hopkins2023, Feldman2023}. Such inhomogeneities, typically washed out in global averages, may create localized sites of massive-star formation that are disproportionately important for producing \ac{GW} progenitors. Extending this work to characterize the host environments of isolated \ac{GW} sources, including the masses, types, and internal regions of galaxies contributing most to progenitor formation, will also enable future multi-messenger applications.

\subsubsection{Population synthesis models}
\label{sec:popsynth}

We use a single binary evolution model from \texttt{COMPAS} in this study, as binary evolution assumptions can contribute uncertainty to the resulting population. Using the same binary evolution model across our populations allows us isolate the effects of varying \SFRD on the \ac{BBH} merger population.
Processes such as common envelope evolution can alter merger rates by more than an order of magnitude across models \citep[e.g.][]{Olejak2021, Broekgaarden2022, Broekgaarden2026}, with uncertainties increasing at the highest \ac{BBH} masses \citep[e.g.][]{Romagnolo2023}.
While there is active research ongoing to use detailed binary models to refine prescriptions \citep[e.g.][]{Marchant2021, Romagnolo2025}, rapid population synthesis codes necessarily rely on simplified assumptions to remain computationally tractable.
Both \citet{Mapelli2017} and \citet{Sgalletta2024} find that binary evolution uncertainties often dominate over \SFRD uncertainties in their impact on \ac{BBH} merger rates.
The \ac{BBH} rates may be overestimated due to the large \ac{BBH} formation efficiencies in population synthesis codes, where 1 in 8 binaries that have a high enough mass to form a \ac{BBH} will form a merging \ac{BBH} \citep{vanSon2025}. Additionally, the authors suggest that \ac{BBH} formation efficiency may not be nearly as strongly metallicity-dependent as shown by previous works, due to updated wind prescriptions, and the degree of importance of metallicity may be formation channel dependent. 
Future work should further explore \SFRD and binary modeling effects for other binary population synthesis codes, formation channels, and model variations.

\subsubsection{Comparison with observed mass distributions}
\label{sec:obscomparison}

Numerous population synthesis studies find differences in the shape and normalization of the \ac{BBH} primary mass distribution as compared to the LVK results \citep[e.g.,][]{Mapelli2017, Neijssel2019, Broekgaarden2022, Mapelli2019, Olejak2021, Iorio2023, Marinacci2025, Smith2026}. This is due to the complex dependence of \ac{BBH} population properties on both binary evolution and \SFRD, both of which remain uncertain.
\citet{MandelBroekgaarden2022} showed that population synthesis studies have produced a span of local merger rates that are below, consistent with, or above the observed range. In particular, common envelope modeling assumptions have been shown to vary significantly across population synthesis models and codes \citep{Broekgaarden2026} and produce a wide range of local merger rates depending on modeling assumptions. Differing prescriptions across different binary evolution codes also significantly impact the form of the primary mass distribution and the contributions of even the same formation channels to the total population \citep{Smith2026}. 
This reveals a complex landscape where the selection of the binary population synthesis code used, parameter choices within a given code, and \SFRD assumptions (Section~\ref{sec:comparisontoearlierwork}) can significantly change the shape and normalization of the primary mass distribution. Our results highlight that \SFRD assumptions can impact the both the normalization of the mass distribution and the contribution of different formation channels at a given primary mass, therefore also impacting its shape (Section~\ref{sec:BBH masses}).

Additionally, the observed mass distribution may be probing other formation channels, in addition to the isolated binary evolution channel considered in this work. In particular, a number of studies have shown that the high mass end of the primary mass distribution may originate from hierarchical mergers in dynamical environments \citep[e.g.,][and references therein]{Antonini2019, Mapelli2021, Mai2026, Ye2026, Vijaykumar2026}. This may explain our models underpredicting the rate at $M_\mathrm{BH, 1} \gtrsim 35$. Modeling these additional channels may be required to reproduce the observed mass distribution, which is beyond the scope of this work.

\subsection{Effects from fit optimization}
\label{sec:optimization}

In addition to effects arising from numerical effects in cosmological simulations and binary population modeling, the optimization of the analytical fit to \SFRD (Section~\ref{sec:model}) can also have a significant impact. 
The choice of optimization strategy affects the quality of the fit and, therefore, the accuracy with which the analytical fit parameters represent \SFRD for \ac{BBH} population modeling. 
Despite the use of the cost function and penalization factor described in Section~\ref{sec:model}, the analytical fit performs poorly at high redshift (Figure~\ref{fig:sfr_z} and \ref{fig:SFRD_Z_z_tng100}) and at both low and high metallicities (Figure~\ref{fig:sfr_Z} and \ref{fig:SFRD_Z_z_tng100}), where \SFRD is intrinsically low. 
In particular,  the analytical fit fails to capture the bimodality of $\mathcal{S}(Z)$ and the truncation of the low-metallicity tail; features that both affect the resulting \ac{BBH} mass distribution (see Section~\ref{sec:BBH masses Z z}).  
While these limitations are inherent to the functional form of the analytical fit,  improvements to the optimization strategy should nonetheless be explored to improve the accuracy of reproducing the \SFRD using an analytical fit.
The choice of optimization algorithm should also be investigated further in the future, in order to identify why the {\tt BFGS} optimization algorithm failed to converge for TNG300-1, potentially due to the bimodality of $\mathcal{S}(Z)$. 



\section{Conclusions}
\label{sec:conclusions}

We model the \ac{BBH} merger population using both the star formation histories of the three IllustrisTNG cosmological simulations of varying mass resolution and simulation box volume (TNG50-1, TNG100-1, and TNG300-1) and analytical fits to each simulation in order to explore the effectiveness of analytical models in representing a more detailed, underlying \SFRD when modeling \ac{BBH} populations. All populations are modeled using the same {\tt COMPAS} binary population synthesis simulations from \citet{vanSon2022a} and \citet{vanSon2022b} to eliminate uncertainties from binary modeling, which we then combine with \SFRD using the {\tt COMPAS} ``cosmic integration'' post-processing tool. To do this, we develop a new method of performing cosmic integration using \SFRD as a 2-D (rather than separate $\mathcal{S}(z)$ and metallicity distribution), such as that which can be derived from a cosmological simulation or observational data. We test whether using an analytical form that combines the Madau \& Dickinson form of $\mathcal{S}(z)$ with a skewed log-normal metallicity distribution---which allows for a low-metallicity tail---is a reliable approximation for the full \SFRD of each TNG simulation when considering the impact on the resulting \ac{BBH} population properties.

We find that the analytical \SFRD model fails to accurately represent \SFRD of the TNG simulations, particularly at high metallicities across redshift and at low metallicities and high redshifts. These discrepancies primarily arise from the poor fit to the metallicity distribution.
The fit is unable to capture the shape of \SFRD at $Z \gtrsim Z_\odot$, missing a secondary bump in the metallicity distribution near $3 Z_\odot$. As the fit attempts to smooth over this feature, it falsely locates the global maximum of the metallicity distribution at a higher metallicity than in the simulation \SFRD. This produces a nontrivial effect on the resulting \ac{BBH} merger population. When using the analytical fit to model \SFRD:
\begin{enumerate}
    \item The high-redshift ($z \gtrsim 4$) \ac{BBH} merger rate is overestimated by up to a factor of $10-10^4$, depending on TNG simulation. This effect becomes more prominent with increasing redshift and decreasing TNG resolution/increasing box size. 
    \item With decreasing resolution/increasing box size, the merger rate peaks at higher redshift.
    \item There is an artificial excess $8\,M_\odot$ feature at in the $z_\mathrm{merger} = 0.2$ primary mass distribution, by up to a factor of 7 at $z_\mathrm{merger} \lesssim 2$. This feature may primarily originate from the CE channel.
\end{enumerate}
However, the locations of features in the primary mass distribution remain consistent across redshift and for both the fit and simulation-based \SFRD for all TNG simulations. 

Additionally, the analytical fit fails to match the low-metallicity \SFRD in a complex way, variable depending on the TNG simulation. For TNG100-1 and TNG300-1, the fit underestimates the high-$z$, low-$Z$ tail of the metallicity distribution by up to an order of magnitude. For TNG50-1, the fit overestimates \SFRD by about a factor of 2. This can affect the modeling of populations at high redshifts and low metallicities ($Z/Z_\odot \lesssim 10^{-2}$). When using the analytical fit to model \SFRD, the high mass rate ($M_1 \gtrsim 20\,M_\odot$ ) is increasingly suppressed with increasing primary mass at $z_\mathrm{merger} = 0.2$. This includes the flattening of the peaks near $33\,M_\odot$ and $45\,M_\odot$. As we do not model binaries with $Z < 10^{-4}$ ($Z/Z_\odot < 7 \times 10^{-3}$), these effects are less prominent in our \ac{BBH} populations than those from the high metallicity \SFRD fit.

Varying the choice of cosmological simulation used to model \SFRD primarily impacts the normalization of the resulting \ac{BBH} population properties, though the analytical model for TNG50-1 (high resolution/small volume) most accurately represents the shape of its \SFRD. At lower resolution and large volume (TNG300-1), the rates are lower and the disagreement between analytical fit and the simulation \SFRD is greatest. We will systematically explore the impact of cosmological mass resolution and box volume on \acp{BBH} populations in future work.

Overall, our results indicate that analytical fits may introduce significant bias to population models, affecting the observables such as the primary mass distribution. Future work should determine whether the artificial features in the primary mass distributions modeled using analytical fits in our work are specific to using \SFRD of the IllustrisTNG simulations. If other cosmological simulations share similar features, it may indicate that analytical fits should be updated or that we should move towards using detailed star formation histories as opposed to analytical models. Observationally-derived star formation histories should also be considered to determine which cosmological simulations are more reliable for \SFRD modeling.
Nevertheless, our results emphasize that the \ac{BBH} mass distribution is sensitive to \SFRD and metallicity assumptions, highlighting that it is critical to improve \SFRD modeling going forward. Ensuring that the star formation history models used alongside population synthesis models are accurate representations of star formation across cosmic time and metallicity can reduce this source of uncertainty in our population models.

As upcoming observing runs and next-generation gravitational-wave detectors are expected to detect stellar-mass \ac{BBH} mergers to increasingly larger redshifts, the comparison of the observed population with models produced using different star formation history assumptions may be able to provide greater constraints on realistic star formation histories based on the redshift evolution of population properties. However, we stress the need to be cautious when using simple analytical fits to represent more complex underlying data, such as those from cosmological simulations. As analytical models smooth over features that cannot be captured with their limited number of parameters, we may lose valuable information regarding the underlying physics and impact further analyses that are based upon those simplified models.

\section{Acknowledgments}

E.B. and S.L. are supported by NSF Grants No.~AST-2307146, No.~PHY-2513337, No.~PHY-090003, and No.~PHY-20043, by NASA Grant No.~21-ATP21-0010, by John Templeton Foundation Grant No.~62840, by the Simons Foundation [MPS-SIP-00001698, E.B.], by the Simons Foundation International [SFI-MPS-BH-00012593-02], and by Italian Ministry of Foreign Affairs and International Cooperation Grant No.~PGR01167. S.L. was additionally supported by the William H. Miller III Graduate Fellowship. 
L.v.S. acknowledges support from the Dutch Research Council (NWO) through the NWO Talent Programme (Veni, Grant DOI: \url{https://doi.org/10.61686/XVIAV86753}).

This work was carried out at the Advanced Research Computing at Hopkins (ARCH) core facility (\url{https://www.arch.jhu.edu/}), which is supported by the NSF Grant No.~OAC-1920103. A.R. acknowledges financial support from the European Research Council for the ERC Consolidator grant DEMOBLACK, under contract no. 770017 and from the German Excellence Strategy via the Heidelberg Cluster of Excellence (EXC 2181 - 390900948) STRUCTURES.

This work made use of the following software packages: \texttt{astropy} \citep{astropy:2013,astropy:2018,astropy:2022}, \texttt{Jupyter} \citep{2007CSE.....9c..21P,kluyver2016jupyter}, \texttt{matplotlib} \citep{Hunter:2007}, \texttt{numpy} \citep{numpy}, \texttt{python} \citep{python}, \texttt{scipy} \citep{2020SciPy-NMeth,scipy_6560517}, \texttt{Cython} \citep{cython:2011}, \texttt{h5py} \citep{collette_python_hdf5_2014,h5py_7560547}, and \texttt{seaborn} \citep{Waskom2021}.

Simulations in this paper made use of the \texttt{COMPAS} rapid binary population synthesis code (version 02.26.03, which is freely available at \url{http://github.com/TeamCOMPAS/COMPAS} \citep{COMPAS}. Model of gravitational wave selection effects based on \citet{Barrett2018}. The STROOPWAFEL adaptive importance sampling algorithm is from \citet{Broekgaarden2019}. Integration over cosmic star formation history outlined in \citep{Neijssel2019}. COMPAS' model of (pulsational) pair instability supernova is from \citet{Stevenson2019}. Pulsar spin and magnetic field evolution is from \citet{Chattopadhyay2020}. Model of chemically homogeneous evolution is from \citet{Riley2021}.

Software citation information aggregated using \texttt{\href{https://www.tomwagg.com/software-citation-station/}{The Software Citation Station}} \citep{software-citation-station-paper,software-citation-station-zenodo}.

\bibliography{myreferences.bib}

\appendix

\section{Impact of axis scaling on interpretation of the fit to $\mathcal{S}(Z)$}
\label{sec:3panelplots}

\begin{figure*}
\centering
    \includegraphics[width=0.8\textwidth,angle=0]{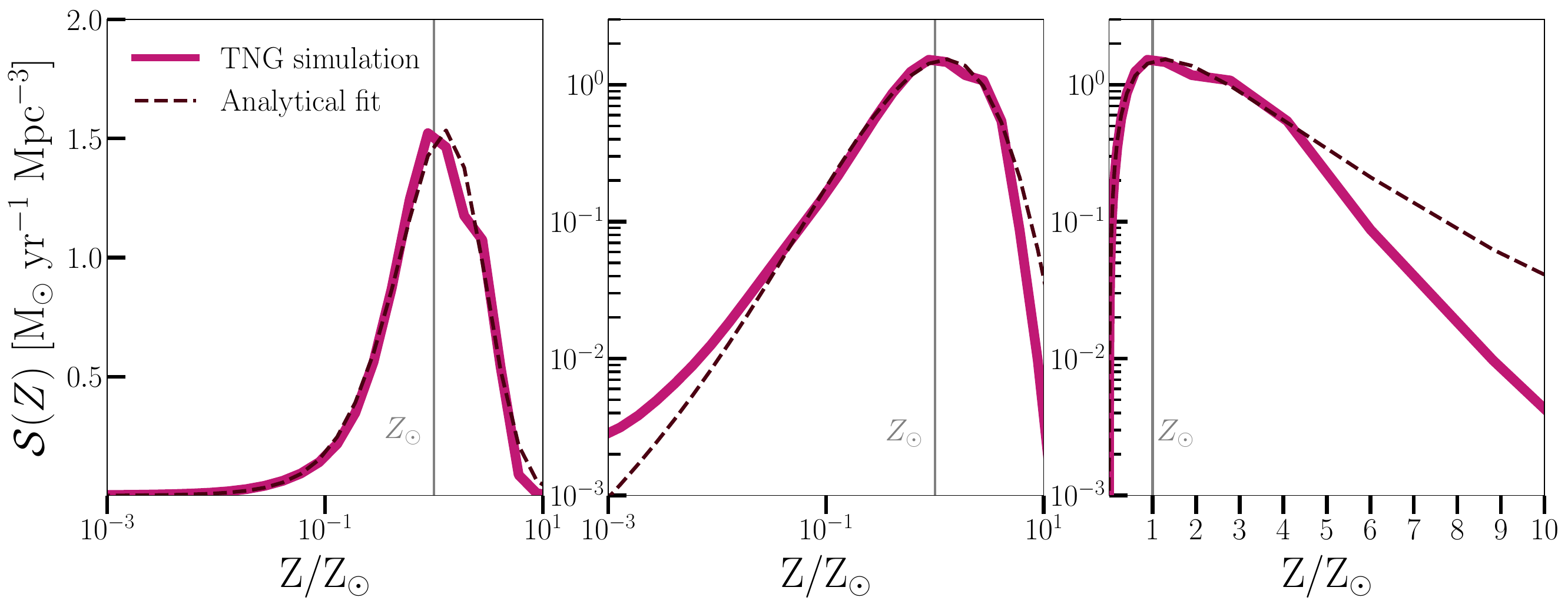}
    \caption{The metallicity distribution $\mathcal{S}(Z)$ for TNG100-1. The solid magenta line is the simulation-based $\mathcal{S}(Z)$ and the dashed line is the fit-based $\mathcal{S}(Z)$. Left: only the horizontal axis on a logarithmic scale; middle: both axes on a logarithmic scale; right: only vertical axis on a logarithmic scale. Note that the vertical axis limits differ between the left and the other two panels. The vertical gray line indicates solar metallicity, for reference.}
        \label{fig:S(Z)_3panels}
\end{figure*}

In Section~\ref{sec:SFRD results}, $\mathcal{S}(Z)$ is shown using different axis scalings to visually connect Figure~\ref{fig:sfr_Z} and the right panel of Figure~\ref{fig:SFRD_Z_z_tng100}. 
In the former, both axes are logarithmic, whereas in the latter the metallicity distribution is shown on a linear vertical scale with logarithmic metallicity.  
These different representations lead to apparent differences in the shape of $\mathcal{S}(Z)$, which can obscure specific aspects of the fit depending on the metallicity range of interest.
To clarify how the choice of axis scaling affects the interpretation of $\mathcal{S}(Z)$, Figure~\ref{fig:S(Z)_3panels} presents the metallicity distribution using three different combinations of linear and logarithmic axes. When $\mathcal{S}(Z)$ is plotted linearly against $\log(Z/Z\odot)$ (right panel, analogous to the right panel of Figure~\ref{fig:SFRD_Z_z_tng100}), the behavior of the fit near the peak of the metallicity distribution is emphasized.
In contrast, using logarithmic scaling on both axes (middle panel, analogous to Figure~\ref{fig:sfr_Z}) highlights deviations at low and high metallicities, where the values of \SFRD are small but important for merging \ac{BBH} formation. 
 Finally, plotting metallicity linearly while scaling $\mathcal{S}(Z)$ logarithmically (left panel) further emphasizes differences near the peak and at super-solar metallicities.
We note that \texttt{COMPAS} does not model binaries  with metallicities above $Z > 0.03$,  corresponding to $Z/Z_\odot \gtrsim 2.11$, which limits the interpretation of the distribution at the highest metallicities. 

\section{Comparison of fit-based and simulation-based mass distributions}
\label{sec:difference}

\begin{figure*}
\centering
    \includegraphics[width=0.8\textwidth,angle=0]{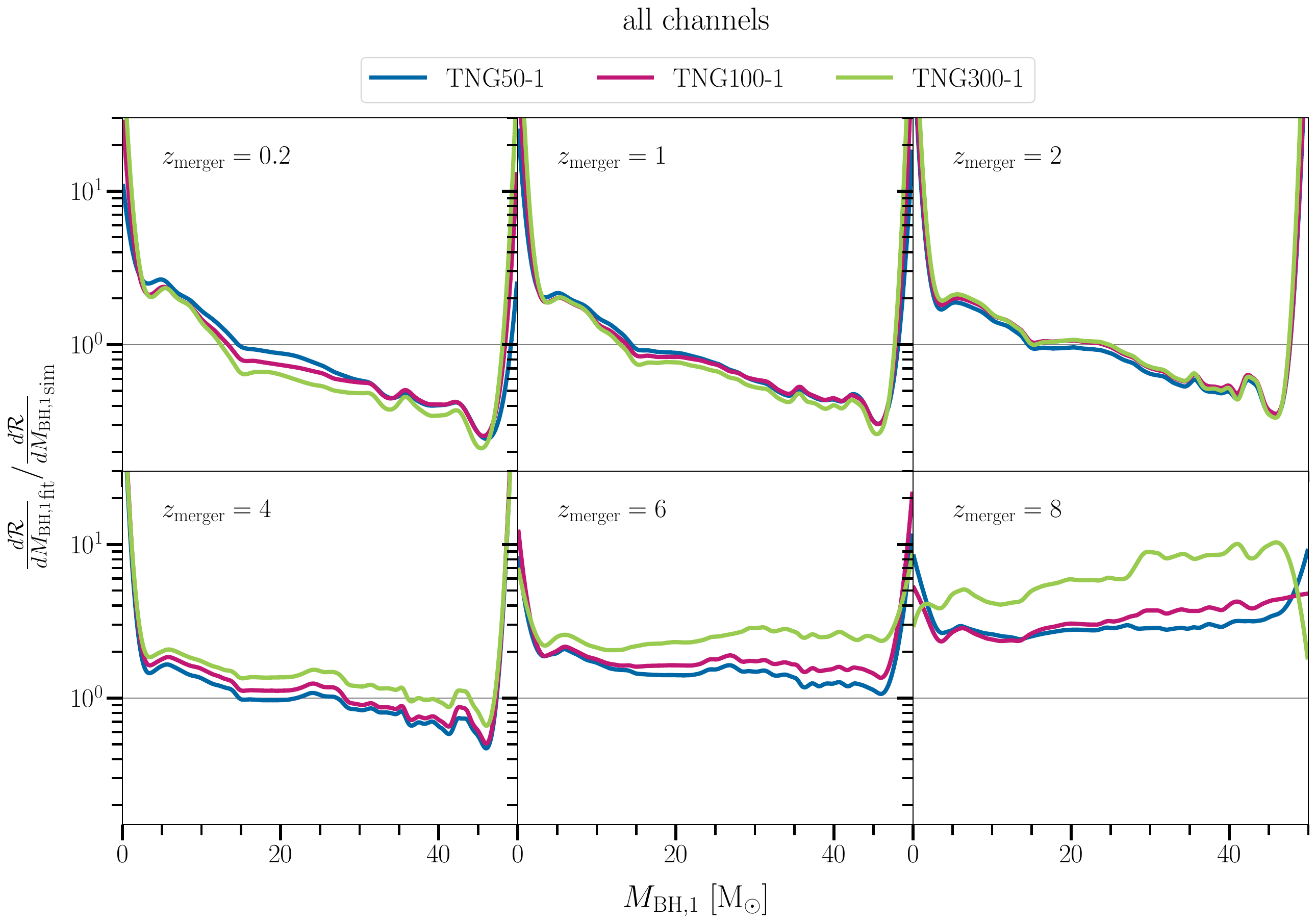}
    \caption{The ratio between the fitted \ac{BBH} primary mass distributions and those from the TNG simulations for $z_\mathrm{merger}=0.2, 1, 2, 4, 6, 8$. Note that the ratio at the edges should not be trusted due to the boundaries of the mass distributions.}
    \label{fig:fractional error mass dist}
\end{figure*}

\begin{figure*}
\centering
    \includegraphics[width=0.8\textwidth,angle=0]{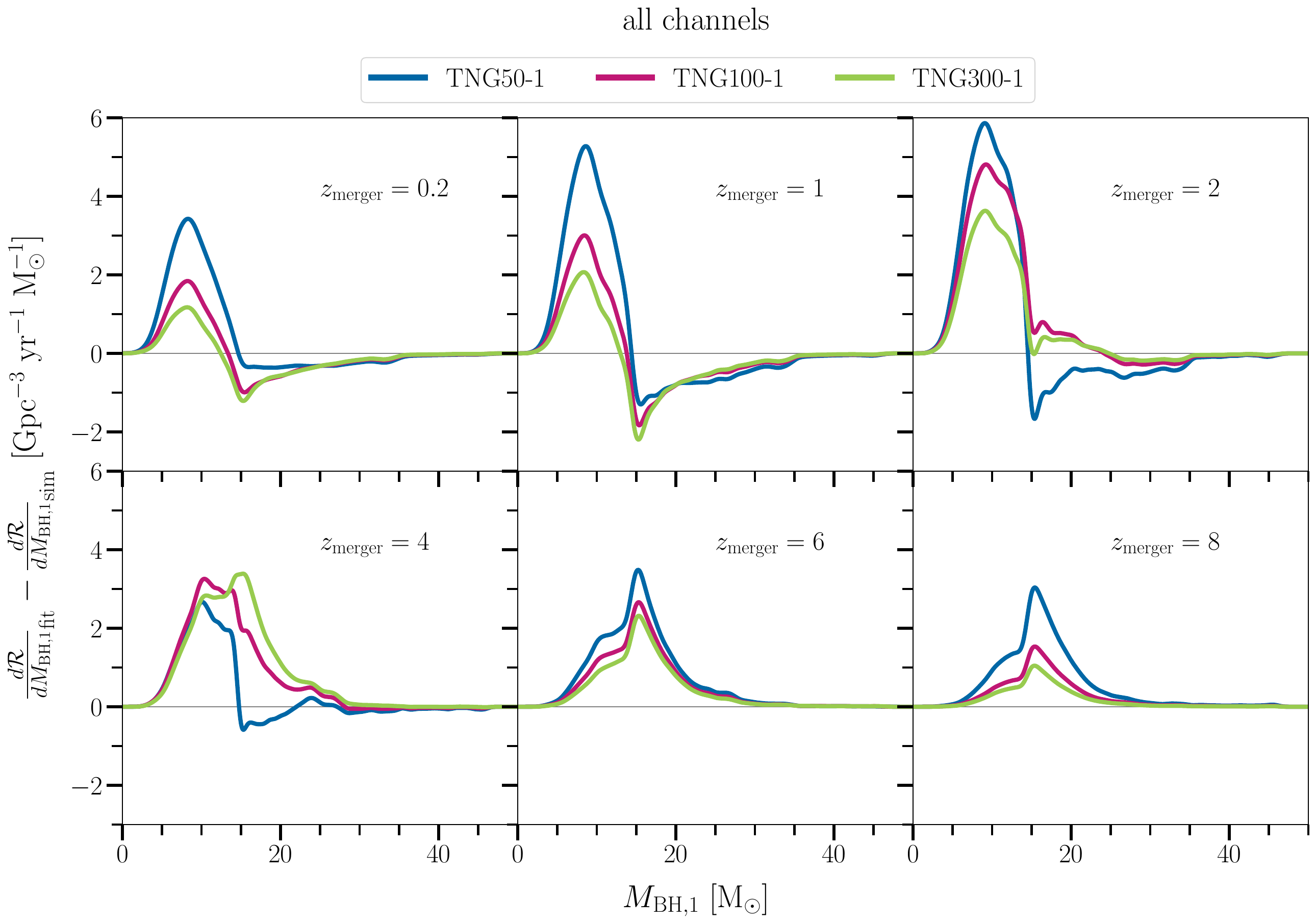}
    \caption{Difference between the \ac{BBH} primary-mass distributions obtained using the analytical fit \SFRD and the TNG simulation \SFRD, shown for $z_\mathrm{merger}=0.2, 1, 2, 4, 6, 8$.}
        \label{fig:massdist difference}
\end{figure*}

To identify which binary systems are misrepresented by the analytical \SFRD fit, we examine the mass distribution of the excess \ac{BBH} mergers that are over-predicted by the analytical \SFRD relative to the TNG simulation \SFRD. Figure~\ref{fig:fractional error mass dist} shows the ratio between the primary mass distribution modeled using the TNG simulation and fit for each TNG simulation in Figure~\ref{fig:massdist comparison}.
The fractional difference has a nearly identical form and normalization across all TNG simulations at $z_\mathrm{merger}=2$, but deviates towards lower and higher redshift. 
This aligns with the expectation that the fit to \SFRD of each TNG performs best near the peak of star formation at $z\approx 2$ (Section~\ref{sec:model}).
At low redshifts ($z_\mathrm{merger}\lesssim 2$, the fit overpredicts the mass distribution at low masses by a factor of 2-3 and underpredicts it at high masses by a factor of 3-5 for all TNGs. 
The ratio between the fit and simulation mass distributions is greatest for TNG300-1 of the three simulations considered, at $z \lesssim 1$ and $z\gtrsim 4$.
At $z\gtrsim 6$, the fit overpredicts the rate at at masses, for all TNG simulations. The fit to TNG300-1 overpredicts the $z_\mathrm{merger} =8$ mass distribution by up by a factor of 10, while the fits to TNG50-1 and TNG100-1 overpredict it by up to a factor of 2 and 4, respectively. This is expected based on the redshift evolution of the merger rate (Figure~\ref{fig:merger rates}).
This shows that the choice of TNG simulation does not have a significant effect on the shape of the mass distribution ratio. The fractional difference flattens with increasing redshift. This flattening indicates that the fit reproduces the shape of the simulation mass distribution, but not its normalization. This may be caused by the poor fit behavior at high redshift and low metallicity, discussed in Section~\ref{sec:SFRD results}, where the fit finds a steeper decline in star formation rate towards high redshift and low metallicities. 

In Figure~\ref{fig:massdist difference}, we highlight the redshift dependence of the excess structure near $M_{\mathrm{BH},1}\simeq8 \Msun$ that arises in the mass distribution when the fit is used.
This behavior explains the apparent disappearance of the $\sim8\,\Msun$ feature at $z\gtrsim2$ in the total mass distributions (Figure~\ref{fig:massdist comparison}), while demonstrating that a residual excess persists at $\sim8\,\Msun$ across all redshifts, alongside a secondary excess near the peak $M_{\mathrm{BH},1}\simeq16\,\Msun$ at $z\gtrsim2$.

At low redshift ($z\lesssim2$), the difference  between the fit and simulation mass distributions peaks near $M_{\mathrm{BH},1}\simeq8 \Msun$. 
As redshift increases, the dominant peak shifts toward higher primary masses, reaching $\sim 16 \Msun$, while a smaller bump near $8\Msun$ remains. 
For TNG50-1, the overproduction at $z_\mathrm{merger}=0.2$ reaches $\sim4~{\rm Gpc}^{-3}{\rm yr}^{-1}\Msun^{-1}$ near $8\Msun$, increasing to approximately $\sim6~{\rm Gpc}^{-3}{\rm yr}^{-1}\Msun^{-1}$ at $z_\mathrm{merger}=2$. 
At higher redshifts, the peak difference decreases and levels off to $\sim 3~{\rm Gpc}^{-3}{\rm yr}^{-1}\Msun^{-1}$. 
At higher primary masses ($M_\mathrm{BH, 1} \gtrsim 16\Msun$), the analytical fit instead underpredicts the merger rate by up to $1$--$2~{\rm Gpc}^{-3}{\rm yr}^{-1}\Msun^{-1}$, with the largest discrepancy occurring around $z_\mathrm{merger}=1$. 

Finally, the magnitude of the excess shows a clear dependence on resolution. 
At $M_{\mathrm{BH},1}\simeq8\,\Msun$, the difference between the fit-based and simulation-based mass distributions in TNG50-1 is approximately a factor of 2 larger than in TNG-100-1, and $\sim3-4$ times larger than in TNG300-1. 
This trend indicates that resolution-dependent features in the simulated \SFRD directly propagate into the predicted \ac{BBH} mass distributions when using analytical fits.

\section{Formation redshift and metallicity of locally merging BBHs}
\label{sec:2dplotregions}

\begin{figure*}
    \centering
    \includegraphics[width=0.8\textwidth,angle=0]{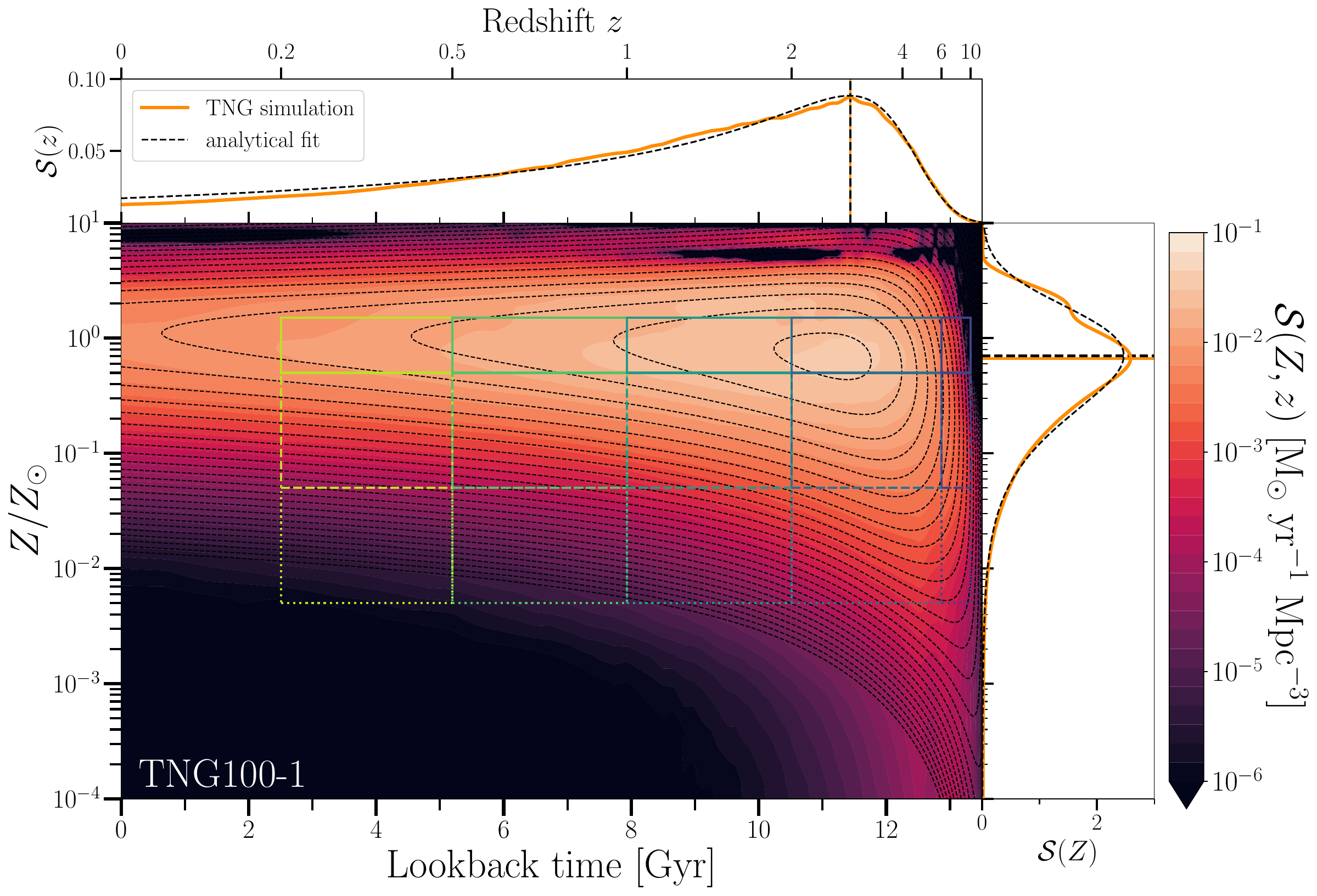}
    \caption{Analogous to Figure~\ref{fig:SFRD_Z_z_tng100}, but with the metallicity and redshift bins used in Figure~\ref{fig:massdist_Z_zform} plotted as rectangles. Note that the maximum \texttt{COMPAS} metallicity is $Z = 0.03$.}
    \label{fig:SFRD_Z_z_tng100_regions}
\end{figure*}

We identify the regions of \SFRD where \acp{BBH} in the $z_\mathrm{merger}=0.2$ primary mass distribution of TNG100-1 originate in order to identify the regimes in which the discrepancies between the fit and simulation star formation histories impact the resulting population (Figure~\ref{fig:SFRD_Z_z_tng100_regions}). We use the formation redshift and metallicity bins used in Figures~\ref{fig:massdist_Z_zform} and \ref{fig:massdist_Z_zform_error}.
We note that the relevant parameter space of \SFRD for locally merging \acp{BBH} includes the ill-fitted region between the metallicity distribution peaks in the highest metallicity bin.
This region preferentially produces lower-mass primary BHs ($M_\mathrm{BH,1} \lesssim 20 \Msun$ that we identify to cause the excess at $8 \Msun$ (Section~\ref{sec:BBH masses Z z}).

\end{document}